\documentclass[12pt]{article}
\usepackage{a4wide}
\usepackage{latexsym}
\usepackage{amsmath}
\usepackage{amsfonts}
\usepackage{amscd}
\usepackage{cite}
\usepackage{graphicx}
\usepackage{float}

\usepackage{pslatex}
\usepackage[latin1]{inputenc}
\usepackage[T1]{fontenc}

\def\bq{\begin{eqnarray}}
\def\eq{\end{eqnarray}}

\def\myplus{+}
\def\myminus{-}
\def\mylbracket{\langle}
\def\myrbracket{\rangle}
\def\mynosign{}

\begin{document}

\thispagestyle{empty}

\begin{flushright}
  MZ-TH/10-27
\end{flushright}

\vspace{1.5cm}

\begin{center}
  {\Large\bf The MHV Lagrangian for a spontaneously broken gauge theory\\
  }
  \vspace{1cm}
  {\large Sebastian Buchta and Stefan Weinzierl\\
\vspace{2mm}
      {\small \em Institut f{\"u}r Physik, Universit{\"a}t Mainz,}\\
      {\small \em D - 55099 Mainz, Germany}\\
  } 
\end{center}

\vspace{2cm}

\begin{abstract}\noindent
  {
Starting from the standard Lagrangian for a $SU(2) \times U(1)$ gauge theory plus a Higgs field
we derive the corresponding ``maximal helicity violating'' (MHV) Lagrangian.
From this MHV Lagrangian one deduces simple diagrammatic rules for the calculation of multi-particle scattering amplitudes.
We arrive at the MHV Lagrangian by a canonical change of the field variables in the light-cone gauge.
We comment on the modifications which occur in a spontaneously broken gauge theory as compared to a pure (unbroken) 
Yang-Mills theory. 
   }
\end{abstract}

\vspace*{\fill}

\newpage

\section{Introduction}

The efficient calculation of scattering amplitudes with many external legs
is a challenging task and needed for phenomenological studies at TeV colliders.
Of particular interest are processes which involve electro-weak gauge bosons.
These processes often lead to the same signatures in the detector as signals of new physics.

In the past years, various new methods for efficient calculations in a gauge theory have
been introduced, motivated by the relation of gluon amplitudes to twistor string
theory~\cite{Witten:2003nn}.  
In particular these methods include the diagrammatic rules of
Cachazo, Svr\v{c}ek and Witten (CSW)~\cite{Cachazo:2004kj}, where tree level QCD
amplitudes are constructed from vertices that are off-shell continuations of
maximal helicity violating (MHV) amplitudes~\cite{Parke:1986gb}, and the recursion relations of
Britto, Cachazo, Feng and Witten~(BCFW)~\cite{Britto:2004ap,Britto:2005fq} that construct
scattering amplitudes from on-shell amplitudes with external momenta shifted
into the complex plane.  
These methods have found numerous applications in tree
level~\cite{Luo:2005rx,Luo:2005my,Britto:2005dg,Badger:2005zh,Forde:2005ue,Quigley:2005cu,Risager:2005vk,Draggiotis:2005wq,Vaman:2005dt,Ozeren:2006ft,Schwinn:2005pi,Schwinn:2006ca,Dinsdale:2006sq,Duhr:2006iq,Draggiotis:2006er,deFlorian:2006ek,deFlorian:2006vu,Rodrigo:2005eu,Ferrario:2006np,Schwinn:2007ee,Hall:2007mz,Schwinn:2008fm}
and
one-loop~\cite{Bidder:2004tx,Bidder:2004vx,Bidder:2005ri,Bedford:2004nh,Britto:2005ha,Bern:2005hs,Bern:2005ji,Bern:2005cq,Bern:2005hh,Forde:2005hh,Berger:2006ci,Berger:2006vq,Berger:2006sh,Britto:2006sj,Xiao:2006vr,Su:2006vs,Xiao:2006vt,Binoth:2006hk,Binoth:2007ca,Ossola:2006us,Ossola:2007bb,Anastasiou:2006jv,Anastasiou:2006gt,Mastrolia:2006ki,Britto:2006fc,Badger:2007si,Forde:2007mi,Ossola:2007ax,Britto:2007tt,Kilgore:2007qr}
calculations in QCD.  
The diagrammatic methods have also been applied to include 
additional non-QCD-type particles, like vector bosons or the 
Higgs boson~\cite{Bern:2004ba,Dixon:2004za,Badger:2004ty,Badger:2005jv}.

The BCFW recursion relations have first been proven with the help of Cauchy's theorem and the vanishing of the amplitudes
at infinity~\cite{Britto:2005fq,Draggiotis:2005wq,Schwinn:2007ee}.
From the BCFW recursion relations one can then deduce the MHV rules~\cite{Risager:2005vk}.
Given the simplicity of the MHV rules it is natural to ask if there is a direct way to transform 
the conventional Lagrangian of Yang-Mills theory into an effective Lagrangian such that the MHV rules can be read off
directly from this effective Lagrangian.
This is indeed possible and has been shown for pure Yang-Mills theory with two different approaches.
The first approach makes use of a canonical transformation in the 
field variables~\cite{Gorsky:2005sf,Mansfield:2005yd,Ettle:2006bw,Ettle:2007qc,Ettle:2008ey,Ettle:2008ed}.
In the second approach one starts from an action in 
twistor space~\cite{Mason:2005kn,Mason:2005zm,Boels:2007qn,Boels:2006ir,Boels:2007gv,Boels:2008fc,Jiang:2008xw}.
The action in twistor space has an extended gauge symmetry.
The conventional Lagrangian and the MHV Lagrangian are then obtained from the action in twistor space for different gauge
choices.

The interest in the major part of the literature has been focused up to now on an unbroken gauge theory.
Equipped with the knowledge and experience from the case of an unbroken gauge theory it is then natural
to ask if these methods can be carried over to the case of a spontaneously broken gauge theory.
This is the question which we want to address in this paper.
We start from the conventional Lagrangian for a $SU(2) \times U(1)$ gauge theory plus a Higgs field 
and derive the corresponding MHV Lagrangian.
From this MHV Lagrangian one obtains simple diagrammatic rules 
for the calculation of scattering amplitudes involving several electro-weak gauge bosons and/or scalar fields.
In this first paper on the MHV formulation of a spontaneously broken gauge theory we try to focus on the essentials.
Therefore we do not include fermions nor do we include QCD.
With the methods presented in this paper the inclusion of these two sectors is in principle straightforward, but leads
to longer formulae.

The motivation for deriving the MHV Lagrangian for a spontaneously broken gauge theory is two-fold:
First of all the diagrammatic rules are helpful in phenomenological applications. Scattering amplitudes with
many external particles involving electro-weak gauge bosons are notoriously cumbersome to calculate with traditional
methods based on Feynman diagrams. The MHV rules offer here an alternative.
Secondly, we are also motivated from a more formal perspective: Reformulating the part of the Lagrangian responsible
for the electro-weak symmetry breaking into a different -- and in certain aspects simpler -- form 
will shed some light on the origin of the symmetry breaking itself.

In order to arrive at the MHV Lagrangian for a spontaneously broken gauge theory we follow the approach based
on a canonical transformation.
On a technical level we profited from the papers by Boels and Schwinn~\cite{Boels:2007pj,Boels:2007qy,Boels:2008ef},
in which they derived the MHV Lagrangian for a pure $U(N)$-Yang-Mills theory plus a massive scalar (without scalar self-interactions).
In these papers the authors treat the mass term for the scalar particle as a perturbation.
This perturbation does not enter
the equation which determines the canonical transformation.
We will proceed similar and treat the Higgs potential as a perturbation.
Our results are also relevant in the case of an unbroken $SU(N)$-gauge theory or an unbroken $SU(N)\times U(1)$-gauge theory
with unequal couplings, both coupled to a scalar field. 
In these cases the canonical transformation induces an additional tower of vertices involving four scalar fields.
In the latter case each vertex of this tower is proportional to the difference of the squares of the couplings 
(and vanishes therefore for a $U(N)$-theory, but not for $SU(N)$ or $SU(N)\times U(1)$).
The inclusion of a $\lambda \Phi^4$-term in the Higgs potential leads straightforwardly to a further tower of vertices with four scalar
fields and proportional to $\lambda$.

In the case of a spontaneously broken gauge theory there are a few additional complications related to
the non-vanishing of the scalar field at infinity and to inverse differential operators. 
We will discuss these in detail in the main part of the paper.
As a final result we find that the MHV formulation of a spontaneously broken gauge theory is the one of an unbroken gauge theory
coupled to a scalar field plus additional towers of vertices all proportional to the vacuum expectation value $v$ of the
scalar field.

This paper is organised as follows:
In the next section we start with a short summary of the notation which we use throughout the paper.
Section~\ref{sect:derivation} is the main part of this article and gives the derivation of the MHV formulation for
a spontaneously broken gauge theory. This section is sub-divided into five steps.
Section~\ref{sect:conclusions} contains a summary and the conclusions.
We have included two appendices:
Appendix~\ref{sect:inverse_differential_operators} is devoted to inverse differential operators.
In Appendix~\ref{sect:solution_coeff_canonical_trafo} we have collected useful information on how the system
of integro-differential equations arising from the canonical transformation is solved.

\section{Notation}
\label{sect:setup}

The derivation of the MHV Lagrangian for the electro-weak theory is simplified by an appropriate notation.
In order to help the reader to follow our arguments in the main part of this article we give in this section
a summary on the notation used throughout this article.

The electro-weak part of the Standard Model is described by a $SU(2) \times U(1)$ gauge theory.
We will denote the gauge fields in the unbroken sector by $W^j_\mu$ (for the $SU(2)$-gauge fields)
and by $B_\mu$ (for the $U(1)$ field).
The conventional Lagrange density for the electro-weak sector is given by
\bq
\label{lagrangian_ew}
{\cal L}_{\mathrm{EW}} & = & 
 - \frac{1}{4} W^j_{\mu\nu} W^{j\;\mu\nu}
 - \frac{1}{4} B_{\mu\nu} B^{\mu\nu}
 + \left( D_\mu \Phi \right)^\dagger \left( D_\mu \Phi \right) + \mu^2 \Phi^\dagger \Phi 
   - \frac{\lambda}{4} \left( \Phi^\dagger \Phi \right)^2,
\eq
where $\Phi$ denotes the Higgs doublet.
The gauge indices of the Higgs doublet are not shown explicitly.
The field strengths are as usual
\bq
 W^j_{\mu\nu} = \partial_\mu W^j_\nu - \partial_\nu W^j_\mu + g f^{jkl} W^k_\mu W^l_\nu,
 & &
 B_{\mu\nu} = \partial_\mu B_\nu - \partial_\nu B_\mu.
\eq
The covariant derivative acting on the Higgs field is given by
\bq
D_\mu & = & \partial_\mu - i g I^j W^j_\mu - i g' \frac{Y}{2} B_\mu.
\eq
$g$ and $g'$ are the couplings of $SU(2)$ and $U(1)$, respectively.
The Higgs doublet has hyper-charge $Y=1$.
The $SU(2)$-matrices are given by $I^j = \frac{1}{2} \sigma^j$, where $\sigma^j$ are the Pauli matrices.
These matrices satisfy
\bq
 \left[ I^j, I^k \right] = i f^{jkl} I^l,
 & & 
 \mbox{Tr}\; I^j I^k = \frac{1}{2} \delta^{jk}.
\eq
It is convenient to introduce a fourth matrix $I^0=\frac{1}{2} {\bf 1}$ and to combine $B_\mu$ and $W^j_\mu$ into
a four-dimensional vector
\bq
\label{four_dim_gauge_vector}
 V^a_\mu & = & \left( B_\mu, W^1_\mu, W^2_\mu, W^3_\mu \right).
\eq
In this paper we use the convention that gauge indices 
from the beginning of the alphabet are in the range $[0,1,2,3]$ and refer to a four-dimensional vector like 
in eq.~(\ref{four_dim_gauge_vector}),
while gauge indices from the middle of the alphabet are in the range $[1,2,3]$ and refer only to the
$SU(2)$ part.

We denote by $A_\mu$, $W^\pm_\mu$ and $Z_\mu$ the eigenstates of the mass matrix. Again we combine them into
a four-dimensional vector
\bq
 X^a_\mu & = & \left( A_\mu, W^+_\mu, W^-_\mu, Z_\mu \right).
\eq
The mass eigenstates $X^a_\mu$ are linear combinations of the states $V^a_\mu$:
\bq
\label{change_to_mass_eigenstates}
 X^a_\mu & = & R^{ab} V^b_\mu.
\eq
The matrix $R^{ab}$ is given by
\bq
\label{rotation_to_mass_eigenstates}
 R^{ab} = \left( \begin{array}{cccc}
          \cos \theta_W & 0 & 0 & -\sin \theta_W \\
          0 & \frac{1}{\sqrt{2}} & -\frac{i}{\sqrt{2}} & 0 \\ 
          0 & \frac{1}{\sqrt{2}} &  \frac{i}{\sqrt{2}} & 0 \\ 
          \sin \theta_W & 0 & 0 & \cos \theta_W \\
          \end{array} \right),
 & &
 \sin \theta_W = \frac{g'}{\sqrt{g^2+{g'}^2}}.
\eq
It is also convenient to introduce the Lie-algebra valued fields
\bq
 {\bf W}_\mu = \frac{g}{i} I^j W^j_\mu,
 & &
 {\bf B}_\mu = \frac{g'}{i} I^0 B_\mu,
\eq
together with the corresponding field strengths
\bq
 {\bf W}_{\mu\nu} = \partial_\mu {\bf W}_\nu - \partial_\nu {\bf W}_\mu + \left[ {\bf W}_\mu, {\bf W}_\nu \right],
 & &
 {\bf B}_{\mu\nu} = \partial_\mu {\bf B}_\nu - \partial_\nu {\bf B}_\mu.
\eq
We will also write
\bq
 {\bf V}_\mu & = & {\bf B}_\mu + {\bf W}_\mu
 =
 \frac{g'}{i} I^0 B_\mu + \frac{g}{i} I^j W^j_\mu.
\eq
With this notation we can write the covariant derivative simply as
\bq
D_\mu & = & \partial_\mu + {\bf W}_\mu + {\bf B}_\mu
 = \partial_\mu + {\bf V}_\mu.
\eq
We will work in light-cone gauge.
We define the light-cone coordinates by
\bq
 x_+=\frac{1}{\sqrt{2}}\left(x_0+x_3\right), 
 \;\;\;
 x_-=\frac{1}{\sqrt{2}}\left(x_0-x_3\right), 
 \;\;\;
 x_{\bot}=\frac{1}{\sqrt{2}}\left(x_1+ix_2\right), 
 \;\;\;
 x_{\bot\ast}=\frac{1}{\sqrt{2}}\left(x_1-ix_2\right).
\eq
With this definition the Minkowski scalar product is given by
\bq
 x^\mu y_\mu & = & 
 x_+ y_- + x_- y_+ - x_{\bot} y_{\bot^\ast} - x_{\bot^\ast} y_{\bot}.
\eq
The contra-variant version of the light-cone coordinates is defined analogously
\bq
 p^+=\frac{1}{\sqrt{2}}\left(p^0+p^3\right), 
 \;\;\;
 p^-=\frac{1}{\sqrt{2}}\left(p^0-p^3\right), 
 \;\;\; 
 p^\bot=\frac{1}{\sqrt{2}}\left(p^1+ip^2\right), 
 \;\;\;
 p^{\bot\ast}=\frac{1}{\sqrt{2}}\left(p^1-ip^2\right).
\eq
Then
\bq
 p^\mu x_\mu & = & p^+ x_+ + p^- x_- + p^\bot x_{\bot\ast} + p^{\bot\ast} x_\bot.
\eq
For the vector $\vec{x}=(x_\myminus,x_{\bot},x_{\bot^\ast})$ we set
\bq
 \vec{p} \cdot \vec{x} & = & p^\myminus x_\myminus + p^\bot x_{\bot\ast} + p^{\bot\ast} x_\bot.
\eq
We define the spinors as
\bq
\label{def_spinors}
\left| p+ \right\rangle = \frac{2^{\frac{1}{4}}}{\sqrt{p^-}} \left( \begin{array}{c}
  p^{\bot^\ast} \\ p^- \end{array} \right),
 & &
\left| p- \right\rangle = \frac{2^{\frac{1}{4}}}{\sqrt{p^-}} \left( \begin{array}{c}
 p^- \\ -p^\bot \end{array} \right),
 \nonumber \\
\left\langle p+ \right| = \frac{2^{\frac{1}{4}}}{\sqrt{p^-}} 
 \left( p^\bot, p^- \right),
 & &
\left\langle p- \right| = \frac{2^{\frac{1}{4}}}{\sqrt{p^-}} 
 \left( p^-, -p^{\bot^\ast} \right).
\eq
This definition applies to all four-vectors $p^\mu$. If the four-vector $p^\mu$ is light-like, the spinors are
the eigenstates of the Dirac equation with eigenvalue zero.
If the four-vector $p^\mu$ is not light-like, eq.~(\ref{def_spinors}) defines the off-shell continuation of the
spinors.
Spinor products are denoted as
\bq
 \langle p q \rangle & = & \langle p - | q + \rangle 
 = \frac{\sqrt{2}}{\sqrt{p^- q^-}} \left( p^- q^{\bot\ast} - q^- p^{\bot\ast} \right),
 \nonumber \\
 \left[ q p \right] & = & 
 \langle q + | p - \rangle = \frac{\sqrt{2}}{\sqrt{p^- q^-}} \left( p^- q^\bot - q^- p^\bot \right).
\eq 
Multiple Fourier integrals will occur frequently and for these integrals we introduce the short-hand notation
\bq
 \int\limits_{(1,...,n)} dP(x) 
 & = &
   \int \frac{d^4p_1}{(2\pi)^4} ... \frac{d^4p_n}{(2\pi)^4}
   e^{-i\left(p_1+...+p_n\right)\cdot x}.
\eq

\section{Derivation of the MHV Lagrangian}
\label{sect:derivation}

In this section we derive the MHV Lagrangian for a spontaneously broken gauge theory,
which is the main result of this paper.
We organise the derivation in five steps.
In the first step we simply choose the light-cone gauge for the $SU(2)$ and the $U(1)$ gauge fields.
In step two we integrate out one component for each gauge field and obtain a Lagrange density which depends
only on the two transverse degrees of freedom for each gauge field.
This Lagrangian is not yet in the MHV form, as it contains both a MHV three-vertex and an anti-MHV three-vertex.
Integrating out one component for each gauge field introduces additional terms which are quartic in the scalar field.
In step three we analyse the vacuum state of the scalar field and expand the scalar field around a minimum of the theory.
In step four we eliminate the anti-MHV three vertex with the help of a canonical transformation.
Finally, in step five we assemble all pieces and give the Lagrangian of a spontaneously broken gauge theory in the MHV form.
 
\subsection{Step 1: Light-cone gauge}

Our starting point is the Lagrangian of the electro-weak sector of the Standard Model
as given in eq.~(\ref{lagrangian_ew}). We can re-write this Lagrangian as
\bq
{\cal L}_{\mathrm{EW}} & = & 
 \frac{1}{2g^2} \; \mbox{Tr} \; {\bf W}_{\mu\nu} {\bf W}^{\mu\nu}
 + \frac{1}{2{g'}^2} \; \mbox{Tr} \; {\bf B}_{\mu\nu} {\bf B}^{\mu\nu}
 + \left( D_\mu \Phi \right)^\dagger \left( D_\mu \Phi \right) + \mu^2 \Phi^\dagger \Phi
   - \frac{\lambda}{4} \left( \Phi^\dagger \Phi \right)^2.
 \;\;\;
\eq
We choose the light-cone gauge
\bq
{\bf W}_\myminus=0,
 & &
{\bf B}_\myminus=0.
\eq
In this gauge we can re-order the Lagrangian as follows:
\bq
 {\cal L}_{\mathrm{EW}} & = & {\cal L}_2 + {\cal L}_3 + {\cal L}_4 + {\cal L}_\Phi + {\cal L}_V,
\eq
such that ${\cal L}_2$ contains all terms bilinear in the gauge fields.
Terms with three or four gauge fields are collected in ${\cal L}_3$ and ${\cal L}_4$, respectively.
${\cal L}_\Phi$ contains the terms bilinear in the scalars as well as couplings of the scalars to the gauge fields.
The Higgs potential is denoted by ${\cal L}_V$.
The explicit expressions read
\bq
\label{lagrangian_light_cone}
 {\cal L}_2 & = & 
 \frac{1}{g^2} 
 \mbox{Tr} \left[ 
                  {\bf W}_\myplus \partial_\myminus^2 {\bf W}_\myplus
                - 2 {\bf W}_\myplus \partial_\myminus \partial_{\bot} {\bf W}_{\bot^\ast}
                - 2 {\bf W}_\myplus \partial_\myminus \partial_{\bot^\ast} {\bf W}_{\bot}
 \right. \nonumber \\
 & & \left.
                + {\bf W}_{\bot} \partial_{\bot^\ast}^2 {\bf W}_{\bot}
                + {\bf W}_{\bot^\ast} \partial_{\bot}^2 {\bf W}_{\bot^\ast}
                + 2 {\bf W}_{\bot^\ast} \left( 2 \partial_\myminus \partial_\myplus - \partial_{\bot} \partial_{\bot^\ast} \right) {\bf W}_{\bot}
           \right]
 \nonumber \\
 & &
 +
 \frac{1}{{g'}^2} 
 \mbox{Tr} \left[ 
                  {\bf B}_\myplus \partial_\myminus^2 {\bf B}_\myplus
                - 2 {\bf B}_\myplus \partial_\myminus \partial_{\bot} {\bf B}_{\bot^\ast}
                - 2 {\bf B}_\myplus \partial_\myminus \partial_{\bot^\ast} {\bf B}_{\bot}
 \right. \nonumber \\
 & & \left.
                + {\bf B}_{\bot} \partial_{\bot^\ast}^2 {\bf B}_{\bot}
                + {\bf B}_{\bot^\ast} \partial_{\bot}^2 {\bf B}_{\bot^\ast}
                + 2 {\bf B}_{\bot^\ast} \left( 2 \partial_\myminus \partial_\myplus - \partial_{\bot} \partial_{\bot^\ast} \right) {\bf B}_{\bot}
           \right],
 \nonumber \\
 {\cal L}_3 & = & 
 \frac{2}{g^2} 
 \mbox{Tr} \left[ 
                 \left( \partial_{\bot} {\bf W}_{\bot^\ast} \right) \left[ {\bf W}_{\bot^\ast}, {\bf W}_{\bot} \right]
               + \left( \partial_{\bot^\ast} {\bf W}_{\bot} \right) \left[ {\bf W}_{\bot}, {\bf W}_{\bot^\ast} \right]
               - \left( \partial_\myminus {\bf W}_{\bot} \right) \left[ {\bf W}_\myplus, {\bf W}_{\bot^\ast} \right]
 \right. \nonumber \\
 & & \left.
               - \left( \partial_\myminus {\bf W}_{\bot^\ast} \right) \left[ {\bf W}_\myplus, {\bf W}_{\bot} \right]
            \right],
 \nonumber 
\eq
\bq
 {\cal L}_4 & = & 
 -\frac{1}{g^2} 
 \mbox{Tr} \left[ {\bf W}_{\bot}, {\bf W}_{\bot^\ast} \right] \left[ {\bf W}_{\bot}, {\bf W}_{\bot^\ast} \right],
 \nonumber \\
 {\cal L}_\Phi & = & - 2 \Phi^\dagger \left( \partial_\myminus \partial_\myplus - \partial_{\bot} \partial_{\bot^\ast} \right) \Phi
 + 
 \Phi^\dagger 
   \left[ 
          \overleftarrow{\partial}_\myminus \left( {\bf W}_\myplus + {\bf B}_\myplus \right) - \left( {\bf W}_\myplus + {\bf B}_\myplus \right) \partial_\myminus
 \right. \nonumber \\
 & & \left.
        - \overleftarrow{\partial}_{\bot} \left( {\bf W}_{\bot^\ast} + {\bf B}_{\bot^\ast} \right) + \left( {\bf W}_{\bot} + {\bf B}_{\bot} \right) \partial_{\bot^\ast}
        - \overleftarrow{\partial}_{\bot^\ast} \left( {\bf W}_{\bot} + {\bf B}_{\bot} \right) + \left( {\bf W}_{\bot^\ast} + {\bf B}_{\bot^\ast} \right) \partial_{\bot}
   \right] \Phi
 \nonumber \\
 & & 
 + 
 \Phi^\dagger 
   \left[ 
        \left( {\bf W}_{\bot^\ast} + {\bf B}_{\bot^\ast} \right) \left( {\bf W}_{\bot} + {\bf B}_{\bot} \right) 
        + \left( {\bf W}_{\bot} + {\bf B}_{\bot} \right) \left( {\bf W}_{\bot^\ast} + {\bf B}_{\bot^\ast} \right) 
   \right] \Phi,
 \nonumber \\
 {\cal L}_V & = & \mu^2 \Phi^\dagger \Phi - \frac{\lambda}{4} \left( \Phi^\dagger \Phi \right)^2.
\eq

\subsection{Step 2: Integrating out ${\bf W}_\myplus$ and ${\bf B}_\myplus$}

We observe that the fields ${\bf W}_\myplus$ and ${\bf B}_\myplus$
occur only quadratically or linearly in eq.~(\ref{lagrangian_light_cone}). 
We can therefore integrate these fields out.
To see how this is done we first consider the case of integrating out a single field $\psi$.
As an example we consider the path integral
\bq
\label{example_subst}
\int {\cal D}\psi \; \exp\int d^4x \; \mbox{Tr}\left( \frac{1}{2} \psi P \psi + \psi K\left(\phi\right) \right).
\eq
We assume that $P$ is a differential operator of even degree and independent of the other fields.
In the case at hand we will have that $P$ is proportional to $\partial_\myminus^2$.
$K(\phi)$ on the other hand may depend on the other fields, which are collectively denoted by $\phi$. 
We would now like to proceed as in the case of an unbroken gauge theory and we would like to make the substitution
\bq
\label{subst_psi}
\psi & \rightarrow & \psi-P^{-1}K\left(\phi\right).
\eq
Here the inverse differential operator $P^{-1}$ appears. 
In the case of a spontaneously broken gauge theory we have to be careful with this inverse differential operator.
Let us first consider the case of an unbroken theory. 
In the appendix~\ref{sect:inverse_differential_operators} we define the space of functions ${\cal F}^{-m,0}$, where
$m$ is a positive integer.
A field belongs to ${\cal F}^{-m,0}$ if the field and its first $m$ inverse derivatives vanish at infinity.
The function spaces ${\cal F}^{-m,0}$ have the property that for sufficiently large $m$ we may use 
partial integration without boundary terms also for the inverse differential operators, 
see eq.~(\ref{partial_integration}) and eq.~(\ref{total_differential}).
The space ${\cal F} = {\cal F}^{-m,0}$ with a suitable $m$ is appropriate for an unbroken gauge theory. 
Within perturbation theory we may assume that all fields lie within this space ${\cal F}$, and that is what is done
in the derivation of the MHV Lagrangian for an unbroken gauge theory.

Now let us turn to the case of a broken gauge theory. We first note that by definition ${\cal F}$ does not include
any function, which does not vanish at infinity. In particular all functions which go to a constant non-zero value at
infinity are not included. This is clearly insufficient for a broken gauge theory.
There the Higgs doublet acquires a vacuum expectation value and goes to a constant at infinity.
Let us therefore denote by ${\cal F}_1$ the space of functions, which consists of ${\cal F}$ and the 
constant functions. If we now consider the differential operator $\partial_\myminus$, we first note that the kernel
of $\partial_\myminus$ are just the functions which are constant in $x_\myminus$.
Therefore we may invert $\partial_\myminus$ on ${\cal F}$, but the application of $\partial_\myminus^{-1}$ on a field of
${\cal F}_1$ is ambiguous.
We may write any field $\phi \in {\cal F}_1$ as the sum of a constant field $\phi_0$ 
and a field $\phi' \in {\cal F}$:
\bq
 \phi(x) & = & \phi_0 + \phi'(x).
\eq
We then set
\bq
 \partial_\myminus^{-1} \phi_0 & = & 0
\eq
and therefore 
\bq
 \partial_\myminus^{-1} \phi(x) & = & \partial_\myminus^{-1} \phi'(x).
\eq
As a consequence we have for all fields $\phi' \in {\cal F}$ the expected relation
\bq
 \partial_\myminus^{-1} \partial_\myminus \phi'(x) 
 & = &
 \partial_\myminus \partial_\myminus^{-1} \phi'(x) 
 =
 \phi'(x),
\eq
but for fields $\phi \in {\cal F}_1$ we have
\bq
 \partial_\myminus^{-1} \partial_\myminus \phi(x) 
 & = &
 \partial_\myminus \partial_\myminus^{-1} \phi(x) 
 =
 \phi'(x).
\eq
With these words of warning we now proceed with the substitution given in eq.~(\ref{subst_psi}).
We anticipate that $K$ may go to a constant $K_0$ at infinity and we write
\bq
 K & = & K_0 + K',
\eq
where $K'$ now falls off at infinity.
We then obtain for the expression in eq.~(\ref{example_subst})
\bq
\int {\cal D}\psi \; \exp\int d^4x \; 
 \mbox{Tr}\left( \frac{1}{2} \psi P \psi + \psi K_0 - \frac{1}{2} K P^{-1} K 
 \right).
\eq
We can neglect the irrelevant factor
\bq
\label{irrelevant}
\int {\cal D}\psi \; \exp\int d^4x \; \mbox{Tr}\left( \frac{1}{2} \psi P \psi + \psi K_0 \right)
\eq
and obtain
\bq
\label{example_integrate_out}
\exp\int d^4x \; \mbox{Tr}\left( - \frac{1}{2} K P^{-1} K 
 \right).
\eq
The result in eq.~(\ref{example_integrate_out}) is identical to the unbroken case, only in eq.~(\ref{irrelevant}) 
we have picked up an extra (irrelevant) term $\mbox{Tr}\; \psi K_0$.
We remark that eq.~(\ref{example_integrate_out}) can equally be written as
\bq
\exp\int d^4x \; \mbox{Tr}\left( - \frac{1}{2} K' P^{-1} K' 
 \right),
\eq
i.e. term proportional to $K_0$ do not contribute. This will be important later, when we expand around the minimum of the
Higgs potential.

Let us now return to ${\bf W}_\myplus$ and ${\bf B}_\myplus$.
For ${\bf W}_\myplus$ we have $P=\frac{2}{g^2} \partial_\myminus^2$
and
\bq
\lefteqn{
 K = } & & \\
 & & \frac{2}{g^2}
 \left\{ -\partial_\myminus \partial_{\bot} {\bf W}_{\bot^\ast} -\partial_\myminus \partial_{\bot^\ast} {\bf W}_{\bot}
 + \left[ \partial_\myminus {\bf W}_{\bot}, {\bf W}_{\bot^\ast} \right]
 + \left[ \partial_\myminus {\bf W}_{\bot^\ast}, {\bf W}_{\bot} \right]
 + g^2 I^j \; \mbox{Tr} \; \left( \Phi \overleftrightarrow{\partial}_\myminus \Phi^\dagger I^j \right)
 \right\},
 \nonumber
\eq
where we used the notation $\Phi \overleftrightarrow{\partial}_\myminus \Phi^\dagger = \Phi \partial_\myminus \Phi^\dagger - \Phi \overleftarrow{\partial}_\myminus \Phi^\dagger$.
For ${\bf B}_\myplus$ we have $P=\frac{2}{{g'}^2} \partial_\myminus^2$
and
\bq
 K & = &
 \frac{2}{{g'}^2}
 \left\{ -\partial_\myminus \partial_{\bot} {\bf B}_{\bot^\ast} -\partial_\myminus \partial_{\bot^\ast} {\bf B}_{\bot}
 + {g'}^2 I^0 \; \mbox{Tr} \left( \Phi \overleftrightarrow{\partial}_\myminus \Phi^\dagger I^0 \right)
 \right\}.
\eq
After integrating out ${\bf W}_\myplus$ and ${\bf B}_\myplus$ we can write the Lagrange density of the electro-weak
sector as
\bq
\label{Lagrangian_transverse}
 {\cal L}_{\mathrm{EW}} & = & 
 {\cal L}_{+-} + {\cal L}_{++-} + {\cal L}_{+--} + {\cal L}_{++--} + {\cal L}_V,
\eq
with
\bq
\label{Lagrangian_transverse2}
 {\cal L}_{+-} & = & 
  \frac{4}{g^2} \mbox{Tr}
  {\bf W}_{\bot^\ast} \left( \partial_\myminus \partial_\myplus - \partial_{\bot} \partial_{\bot^\ast} \right) {\bf W}_{\bot}
+
  \frac{4}{{g'}^2} \mbox{Tr}
  {\bf B}_{\bot^\ast} \left( \partial_\myminus \partial_\myplus - \partial_{\bot} \partial_{\bot^\ast} \right) {\bf B}_{\bot}
 \nonumber \\
 & & 
- 2 \Phi^\dagger \left( \partial_\myminus \partial_\myplus - \partial_{\bot} \partial_{\bot^\ast} \right) \Phi,
 \nonumber \\
 {\cal L}_{++-} & = & 
  \frac{4}{g^2} \mbox{Tr} \;
     \left( \partial_{\bot^\ast} {\bf W}_{\bot} \right) \partial_\myminus^{-1} \left[ {\bf W}_{\bot}, \partial_\myminus {\bf W}_{\bot^\ast} \right]
 \nonumber \\
 & &
 - \mbox{Tr} \; \left[ \left( {\bf W}_{\bot} + {\bf B}_{\bot} \right) \left( \Phi \overleftrightarrow{\partial}_{\bot^\ast} \Phi^\dagger \right) \right]
 +
 \mbox{Tr} \;  
   \left[ 
           \left( {\bf W}_{\bot} + {\bf B}_{\bot}
           \right) \partial_\myminus^{-1} \partial_{\bot^\ast} \left( \Phi \overleftrightarrow{\partial}_\myminus \Phi^\dagger \right)
 \right],
 \nonumber \\
 {\cal L}_{+--} & = & 
  \frac{4}{g^2} \mbox{Tr} \;
     \left( \partial_{\bot} {\bf W}_{\bot^\ast} \right) \partial_\myminus^{-1} \left[ {\bf W}_{\bot^\ast}, \partial_\myminus {\bf W}_{\bot} \right]
 \nonumber \\
 & &
  - \mbox{Tr} \; \left[ \left( {\bf W}_{\bot^\ast} + {\bf B}_{\bot^\ast} \right) \left( \Phi \overleftrightarrow{\partial}_{\bot} \Phi^\dagger \right) \right]
 +
 \mbox{Tr} \;  
   \left[ 
           \left( {\bf W}_{\bot^\ast} + {\bf B}_{\bot^\ast} 
           \right) \partial_\myminus^{-1} \partial_{\bot} \left( \Phi \overleftrightarrow{\partial}_\myminus \Phi^\dagger \right)
 \right],
 \nonumber \\
 {\cal L}_{++--} & = & 
  -\frac{4}{g^2} \mbox{Tr} \;
          \left[ {\bf W}_{\bot^\ast}, \partial_\myminus {\bf W}_{\bot} \right] \partial_\myminus^{-2} \left[ {\bf W}_{\bot}, \partial_\myminus {\bf W}_{\bot^\ast} \right]
 \nonumber \\
 & &
 + 
 \mbox{Tr} \;  
   \left[ 
        \left( {\bf W}_{\bot^\ast} + {\bf B}_{\bot^\ast} \right) \left( {\bf W}_{\bot} + {\bf B}_{\bot} \right) \Phi \Phi^\dagger
        + \left( {\bf W}_{\bot} + {\bf B}_{\bot} \right) \left( {\bf W}_{\bot^\ast} + {\bf B}_{\bot^\ast} \right) \Phi \Phi^\dagger
   \right] 
 \nonumber \\
 & & 
 + 
 \mbox{Tr} \;  
   \left[ 
           \left( \partial_\myminus^{-1} \left[ \partial_\myminus {\bf W}_{\bot}, {\bf W}_{\bot^\ast} \right]
                  + \partial_\myminus^{-1} \left[ \partial_\myminus {\bf W}_{\bot^\ast}, {\bf W}_{\bot} \right]
           \right) \partial_\myminus^{-1} \left( \Phi \overleftrightarrow{\partial}_\myminus \Phi^\dagger \right)
 \right]
 \nonumber \\
 & &
  + 
 \frac{g^2}{4} \mbox{Tr} \; 
  \left[ \partial_\myminus^{-1} \left( \Phi \overleftrightarrow{\partial}_\myminus \Phi^\dagger \right) \right]
  \left[ \partial_\myminus^{-1} \left( \Phi \overleftrightarrow{\partial}_\myminus \Phi^\dagger \right) \right]
 \nonumber \\
 & &
  + 
 \frac{1}{8} \left( {g'}^2 - g^2 \right) 
  \mbox{Tr} \; 
  \left[ \partial_\myminus^{-1} \left( \Phi \overleftrightarrow{\partial}_\myminus \Phi^\dagger \right) \right]
  \mbox{Tr} \; 
  \left[ \partial_\myminus^{-1} \left( \Phi \overleftrightarrow{\partial}_\myminus \Phi^\dagger \right) \right],
\eq
and ${\cal L}_V$ is given as in eq.~(\ref{lagrangian_light_cone}).
The Lagrange density in eq.~(\ref{Lagrangian_transverse}) and eq.~(\ref{Lagrangian_transverse2})
contains now only the transverse degrees of freedom for the fields ${\bf W}$ and ${\bf B}$.
In associating terms with a scalar field $\Phi$ to the individual pieces in eq.~(\ref{Lagrangian_transverse2})
we have counted a field $\Phi$ as ``+'' and a field $\Phi^\dagger$ as ``-''.

\subsection{Step 3: Expansion around the minimum}

We are interested in a spontaneously broken gauge theory.
Up to now we parametrised the fields as in an unbroken gauge theory.
We now expand the fields around a minimum of the theory.
To find the minimum we look at the self-interactions of the scalar field.
If we ignore the gauge fields the Lagrangian reduces to
\bq
\label{Lagrangian_Higgs_new}
 {\cal L}_{\mathrm{Higgs}} & = &
 - \Phi^\dagger(x) \Box \Phi(x)
 + \mu^2 \Phi^\dagger(x) \Phi(x) 
 - \frac{\lambda}{4} \left( \Phi^\dagger(x) \Phi(x) \right)^2
 \nonumber \\
 & &
  + 
 \frac{g^2}{4} \mbox{Tr} \; 
  \left[ \partial_\myminus^{-1} \left( \Phi(x) \overleftrightarrow{\partial}_\myminus \Phi^\dagger(x) \right) \right]
  \left[ \partial_\myminus^{-1} \left( \Phi(x) \overleftrightarrow{\partial}_\myminus \Phi^\dagger(x) \right) \right]
 \nonumber \\
 & &
  + 
 \frac{1}{8} \left( {g'}^2 - g^2 \right) 
  \mbox{Tr} \; 
  \left[ \partial_\myminus^{-1} \left( \Phi(x) \overleftrightarrow{\partial}_\myminus \Phi^\dagger(x) \right) \right]
  \mbox{Tr} \; 
  \left[ \partial_\myminus^{-1} \left( \Phi(x) \overleftrightarrow{\partial}_\myminus \Phi^\dagger(x) \right) \right].
\eq
The first line is just the standard Lagrange density for the Higgs field.
The terms in the second and third line originate from ${\cal L}_{++--}$ in eq.~(\ref{Lagrangian_transverse2}).
These terms are quartic in the scalar fields and involve derivatives.
The attentative reader might now fear that these additional terms modify the position of the minimum, maybe even
in a momentum dependent way.
This is not the case as we will show now.
To find the minimum we 
write the scalar field $\Phi\left(x\right)$ as the sum of a constant field $\Phi_0$ and a new
field $\Phi'\left(x\right)$:
\bq
\label{shift_scalar}
 \Phi\left(x\right) & = & \Phi_0 + \Phi'\left(x\right).
\eq
Inserting this splitting into the Lagrangian of eq.~(\ref{Lagrangian_Higgs_new}) we determine the minimum
(and therefore $\Phi_0$) from the requirement that the terms linear in $\Phi'(x)$ vanish.
Let us first discuss the additional terms in eq.~(\ref{Lagrangian_Higgs_new}).
We examine the combination
$\partial_\myminus^{-1} \left( \Phi(x) \overleftrightarrow{\partial}_\myminus \Phi^\dagger(x) \right)$ and find
\bq
 \partial_\myminus^{-1} 
 \left( \Phi\left(x\right) \overleftrightarrow{\partial}_\myminus \Phi^\dagger\left(x\right) \right) 
 & = &
 \partial_\myminus^{-1} 
 \left( \Phi'\left(x\right) \overleftrightarrow{\partial}_\myminus \Phi'{}^\dagger\left(x\right) \right) 
 + \Phi_0 \Phi'{}^\dagger\left(x\right)
 - \Phi'\left(x\right) \Phi_0^\dagger.
\eq
This combination has a term linear in $\Phi'(x)$ and a term which is quadratic in $\Phi'(x)$.
In the second and third line of eq.~(\ref{Lagrangian_Higgs_new}) this combination occurs squared.
Therefore these terms are at least quadratic in $\Phi'(x)$ and do not contribute to the position of the
minimum.
Therefore the minimum is given as usual by the solution of the equation
\bq
 \mu^2 - \frac{1}{2} \lambda \Phi_0^\dagger \Phi_0 & = & 0.
\eq
We set
\bq
 \Phi_0 & = & \frac{1}{\sqrt{2}} \left( \begin{array}{c} 0 \\ v \\ \end{array} \right),
 \;\;\;\;\;\;
 v = 2 \sqrt{\frac{\mu^2}{\lambda}}.
\eq
The components of the new field $\Phi'(x)$ are written as
\bq
 \Phi'(x) & = & 
 \frac{1}{\sqrt{2}}
 \left( \begin{array}{c}
 \phi_1(x) + i \phi_2(x) \\
 H(x) + i \chi(x) \\
 \end{array} \right).
\eq
Let us examine closer the terms ${\cal L}_{++-}$ and ${\cal L}_{+--}$ in eq.~(\ref{Lagrangian_transverse2}).
These terms are invariant under the shift of the scalar field given in eq.~(\ref{shift_scalar}) as can be seen as follows:
If we look at ${\cal L}_{++-}$ we find that the combination
\bq
\lefteqn{
  \left( \Phi(x) \overleftrightarrow{\partial}_{\bot^\ast} \Phi^\dagger(x) \right) 
  - \partial_\myminus^{-1} \partial_{\bot^\ast} \left( \Phi(x) \overleftrightarrow{\partial}_\myminus \Phi^\dagger(x) \right)
 = }
 & & \nonumber \\
 & &
  \left( \Phi'(x) \overleftrightarrow{\partial}_{\bot^\ast} \Phi'{}^\dagger(x) \right) 
  - \partial_\myminus^{-1} \partial_{\bot^\ast} \left( \Phi'(x) \overleftrightarrow{\partial}_\myminus \Phi'{}^\dagger(x) \right)
\eq
transforms invariantly under a shift of the scalar field. A similar relation holds if we replace $\partial_{\bot^\ast}$ by
$\partial_\bot$, which in turn can be applied to ${\cal L}_{+--}$.

After parametrising the fields around the minimum we can write down the Lagrange density in terms of the new scalar field $\Phi'(x)$.
In order to economise on the notational side 
we relabel the new scalar field $\Phi'(x)$ by $\Phi(x)$.
Ignoring a constant term the Lagrange density is then given by
\bq
\label{Lagrangian_transverse3}
 {\cal L}_{\mathrm{EW}} & = & 
 {\cal L}_{+-} + {\cal L}_{++-} + {\cal L}_{+--} + {\cal L}_{++--} + {\cal L}_V
 \nonumber \\
 & &
 + {\cal L}_{++--}' + {\cal L}_V' + {\cal L}_{++--}'' + {\cal L}_V'',
\eq
where 
${\cal L}_{+-}$, ${\cal L}_{++-}$, ${\cal L}_{+--}$ and ${\cal L}_{++--}$ have been given in eq.~(\ref{Lagrangian_transverse2})
and ${\cal L}_V$ has been given in eq.~(\ref{lagrangian_light_cone}).
The new terms ${\cal L}_{++--}'$ and ${\cal L}_V'$ are proportional to the vacuum expectation value $v$ and given by
\bq
 {\cal L}_{++--}' & = &
 \mbox{Tr} \;  
   \left[ 
        \left( {\bf W}_{\bot^\ast} + {\bf B}_{\bot^\ast} \right) \left( {\bf W}_{\bot} + {\bf B}_{\bot} \right) 
        + \left( {\bf W}_{\bot} + {\bf B}_{\bot} \right) \left( {\bf W}_{\bot^\ast} + {\bf B}_{\bot^\ast} \right) 
   \right] 
           \left( \Phi_0 \Phi^\dagger + \Phi \Phi_0^\dagger \right)
 \nonumber \\
 & & 
 +
 \mbox{Tr} \;  
   \left[ 
           \left( \partial_\myminus^{-1} \left[ \partial_\myminus {\bf W}_{\bot}, {\bf W}_{\bot^\ast} \right]
                  + \partial_\myminus^{-1} \left[ \partial_\myminus {\bf W}_{\bot^\ast}, {\bf W}_{\bot} \right]
           \right) \left( \Phi_0 \Phi^\dagger - \Phi \Phi_0^\dagger \right)
 \right]
 \nonumber \\
 & &
 + 
 \frac{g^2}{2} \mbox{Tr} \; 
  \left( \Phi_0 \Phi^\dagger - \Phi \Phi_0^\dagger \right)
  \left[ \partial_\myminus^{-1} \left( \Phi \overleftrightarrow{\partial}_\myminus \Phi^\dagger \right) \right]
 \nonumber \\
 & &
  + 
 \frac{1}{4} \left( {g'}^2 - g^2 \right) 
  \mbox{Tr} \; 
  \left( \Phi_0 \Phi^\dagger - \Phi \Phi_0^\dagger \right)
  \mbox{Tr} \; 
  \left[ \partial_\myminus^{-1} \left( \Phi \overleftrightarrow{\partial}_\myminus \Phi^\dagger \right) \right],
 \nonumber \\
 {\cal L}_V' & = &
 - \frac{1}{2} \lambda \left( \Phi^\dagger \Phi \right) \left( \Phi_0^\dagger \Phi + \Phi^\dagger \Phi_0 \right).
\eq
The new terms ${\cal L}_{++--}''$ and ${\cal L}_V''$ are proportional to $v^2$ and given by
\bq
 {\cal L}_{++--}'' & = &
 \Phi_0^\dagger
   \left[ 
        \left( {\bf W}_{\bot^\ast} + {\bf B}_{\bot^\ast} \right) \left( {\bf W}_{\bot} + {\bf B}_{\bot} \right) 
        + \left( {\bf W}_{\bot} + {\bf B}_{\bot} \right) \left( {\bf W}_{\bot^\ast} + {\bf B}_{\bot^\ast} \right) 
   \right] 
  \Phi_0 
 \nonumber \\
 & &
 + 
 \frac{g^2}{4} \mbox{Tr} \; 
  \left( \Phi_0 \Phi^\dagger - \Phi \Phi_0^\dagger \right)
  \left( \Phi_0 \Phi^\dagger - \Phi \Phi_0^\dagger \right)
 \nonumber \\
 & &
  + 
 \frac{1}{8} \left( {g'}^2 - g^2 \right) 
  \mbox{Tr} \; 
  \left( \Phi_0 \Phi^\dagger - \Phi \Phi_0^\dagger \right)
  \mbox{Tr} \; 
  \left( \Phi_0 \Phi^\dagger - \Phi \Phi_0^\dagger \right),
 \nonumber \\
 {\cal L}_V'' & = &
 - \frac{1}{4} \lambda 
 \left[ 
        \left( \Phi_0^\dagger \Phi \right)^2
      + \left( \Phi^\dagger \Phi_0 \right)^2
      + 2 \left( \Phi_0^\dagger \Phi \right) \left( \Phi^\dagger \Phi_0 \right) 
      + 2 \left( \Phi^\dagger \Phi \right) \left( \Phi_0^\dagger \Phi_0 \right)
 \right].
\eq

\subsection{Step 4: Canonical transformation}

In the fourth step we eliminate the non-MHV vertices contained in ${\cal L}_{++-}$ by 
a canonical change of the field variables.
This step is similar to what has been done in the case of an unbroken gauge theory.
We can rely on the results obtained for a 
pure gauge theory~\cite{Mansfield:2005yd,Ettle:2006bw,Ettle:2007qc,Ettle:2008ey,Ettle:2008ed}
and for a gauge theory coupled to scalar fields~\cite{Boels:2007pj,Boels:2007qy,Boels:2008ef}.
The only modification which we have to make is to include an additional $U(1)$ field.

To motivate the canonical transformation we treat the variable $x_\myplus$ as a time variable
and collect the remaining three variables in a vector $\vec{x}=(x_\myminus,x_{\bot},x_{\bot^\ast})$. 
In order to simplify the notation we will suppress the dependence of the fields on $x_\myplus$ and write
$\phi(\vec{x})$ instead of $\phi(x_\myplus,\vec{x})$.
We will denote the new fields after the canonical transformation with a tilde, e.g.
\bq
 {\bf W} \rightarrow {\bf \tilde{W}},
 \;\;\;
 {\bf B} \rightarrow {\bf \tilde{B}},
 \;\;\;
 \Phi \rightarrow \tilde{\Phi}.
\eq
Now let us look again at eq.~(\ref{Lagrangian_transverse}) and eq.~(\ref{Lagrangian_transverse2}).
The ``momenta'' conjugate to $W_\bot^j$, $B_\bot$ and $\Phi$ are
\bq
 \frac{\delta {\cal L}_{\mathrm{EW}}}{\delta \partial_\myplus W_{\bot}^j}
 =
 2 \partial_\myminus W_{\bot^\ast}^j,
 \;\;\;\;\;\;
 \frac{\delta {\cal L}_{\mathrm{EW}}}{\delta \partial_\myplus B_{\bot}}
 =
 2 \partial_\myminus B_{\bot^\ast},
 \;\;\;\;\;\;
 \frac{\delta {\cal L}_{\mathrm{EW}}}{\delta \partial_\myplus \Phi}
 =
 2 \partial_\myminus \Phi^\dagger.
\eq
We look for a canonical transformation, where the generating function of  the transformation depends 
on the new ``coordinates'' ${\bf \tilde{W}}_{\bot}$, 
${\bf \tilde{B}}_{\bot}$, $\tilde{\Phi}$
and the old ``momenta'' $\partial_\myminus {\bf W}_{\bot^\ast}$, 
$\partial_\myminus {\bf B}_{\bot^\ast}$, $\partial_\myminus \Phi^\dagger$:
\bq
\lefteqn{
 G\left[{\bf \tilde{W}}_{\bot}, {\bf \tilde{B}}_{\bot}, \tilde{\Phi}, 
        \partial_\myminus {\bf W}_{\bot^\ast}, \partial_\myminus {\bf B}_{\bot^\ast}, \partial_\myminus \Phi^\dagger \right]
 =  
 \int d^3y 
  \left\{
  W^j_{\bot}\left[{\bf \tilde{W}}_{\bot}(\vec{y})\right] \partial_\myminus W_{\bot^\ast}^j(\vec{y})
 \right. 
 } & & \nonumber \\
 & & \left.
  + B_{\bot}\left[{\bf \tilde{B}}_{\bot}(\vec{y})\right] \partial_\myminus B_{\bot^\ast}(\vec{y})
  + \Phi_i\left[\tilde{\Phi}(\vec{y}),{\bf \tilde{W}}_{\bot}(\vec{y}),{\bf \tilde{B}}_{\bot}(\vec{y})\right] \partial_\myminus \Phi_i^\dagger(\vec{y}) \right\}.
\eq
The new ``momenta'' are then given by
\bq
\label{new_momenta}
 \partial_\myminus \tilde{W}_{\bot^\ast}^j(\vec{x})
 & = & \int d^3y \; \frac{\delta W_{\bot}^k(\vec{y})}{\delta \tilde{W}_{\bot}^j(\vec{x})} 
                 \; \partial_\myminus W_{\bot^\ast}^k(\vec{y})
     + \int d^3y \; \frac{\delta \Phi_i(\vec{y})}{\delta \tilde{W}_{\bot}^j(\vec{x})}
                 \; \partial_\myminus \Phi_i^\dagger(\vec{y}),
 \nonumber \\
 \partial_\myminus \tilde{B}_{\bot^\ast}(\vec{x})
 & = & \int d^3y \; \frac{\delta B_{\bot}(\vec{y})}{\delta \tilde{B}_{\bot}(\vec{x})} 
                 \; \partial_\myminus B_{\bot^\ast}(\vec{y})
     + \int d^3y \; \frac{\delta \Phi_i(\vec{y})}{\delta \tilde{B}_{\bot}(\vec{x})}
                 \; \partial_\myminus \Phi_i^\dagger(\vec{y}),
 \nonumber \\
 \partial_\myminus \tilde{\Phi}_{i_1}^\dagger(\vec{x}) 
 & = & \int d^3y \frac{\delta \Phi_{i_2}(\vec{y})}{\delta \tilde{\Phi}_{i_1}(\vec{x})}
                 \; \partial_\myminus \Phi_{i_2}^\dagger(\vec{y}).
\eq
The transformation should eliminate the unwanted ${\cal L}_{++-}$ term, therefore we require
\bq
\label{elimination_++-}
\lefteqn{
 {\cal L}_{+-}\left[{\bf \tilde{W}}_{\bot},{\bf \tilde{W}}_{\bot^\ast},{\bf \tilde{B}}_{\bot},{\bf \tilde{B}}_{\bot^\ast},\tilde{\Phi},\tilde{\Phi}^\dagger\right]
 = } & &
 \nonumber \\
 & &
 {\cal L}_{+-}\left[{\bf W}_{\bot},{\bf W}_{\bot^\ast},{\bf B}_{\bot},{\bf B}_{\bot^\ast},\Phi,\Phi^\dagger\right]
 +
 {\cal L}_{++-}\left[{\bf W}_{\bot},{\bf W}_{\bot^\ast},{\bf B}_{\bot},{\bf B}_{\bot^\ast},\Phi,\Phi^\dagger\right].
\eq
The fact that the transformation is canonical implies
\bq
\label{tranformation_kinetic_terms}
\lefteqn{
 \int d^3x \; 
 \left[
 2 \left( \partial_\myminus W_{\bot^\ast}^j \right) \left( \partial_\myplus W_{\bot}^j \right) 
 +
 2 \left( \partial_\myminus B_{\bot^\ast} \right) \left( \partial_\myplus B_{\bot} \right) 
 +
 2 \left( \partial_\myminus \Phi^\dagger \right) \left( \partial_\myplus \Phi \right) 
 \right]
 = } & & \nonumber \\
 & &
 \int d^3x \; 
 \left[
 2 \left( \partial_\myminus \tilde{W}_{\bot^\ast} \right) \left( \partial_\myplus \tilde{W}_{\bot} \right) 
 +
 2 \left( \partial_\myminus \tilde{B}_{\bot^\ast} \right) \left( \partial_\myplus \tilde{B}_{\bot} \right) 
 +
 2 \left( \partial_\myminus \tilde{\Phi}^\dagger \right) \left( \partial_\myplus \tilde{\Phi} \right)
 \right].
\eq
We then plug the expressions in eq.~(\ref{new_momenta}) into eq.~(\ref{elimination_++-}) and use
eq.~(\ref{tranformation_kinetic_terms}).
It is convenient to introduce the following two differential operators
\bq
 \omega = \frac{\partial_{\bot} \partial_{\bot^\ast}}{\partial_\myminus},
 & &
 \zeta = \frac{\partial_{\bot^\ast}}{\partial_\myminus}.
\eq
From the coefficients of $\partial_\myminus {\bf W}_{\bot^\ast}$, 
$\partial_\myminus {\bf B}_{\bot^\ast}$ and $\partial_\myminus \Phi^\dagger$
we find three integro-differential equations
\bq
\label{eq_canonical_trafo}
\lefteqn{
 \omega B_{\bot}(\vec{x}) 
 =  
 \int d^3y \frac{\delta B_{\bot}(\vec{x})}{\delta \tilde{B}_{\bot}(\vec{y})} \omega_y \tilde{B}_{\bot}(\vec{y}),
} & &
\nonumber \\
\lefteqn{
 \omega W_{\bot}^j(\vec{x}) - g f^{jkl} \left( \zeta W_{\bot}^k(\vec{x}) \right) W_{\bot}^l(\vec{x}) 
 =  
 \int d^3y \frac{\delta W_{\bot}^j(\vec{x})}{\delta \tilde{W}_{\bot}^k(\vec{y})} 
 \omega_y \tilde{W}_{\bot}^k(\vec{y}),
} & &
\nonumber \\
\lefteqn{
 \omega \Phi_{i_1}(\vec{x}) 
 + i \left[ \zeta \left( g I^j_{i_1 i_2} W^j_{\bot}(\vec{x}) + g' I^0_{i_1 i_2} B_{\bot}(\vec{x}) \right) \right] \Phi_{i_2}(\vec{x})
 - i \zeta \left[ \left( g I^j_{i_1 i_2} W^j_{\bot}(\vec{x}) + g' I^0_{i_1 i_2} B_{\bot}(\vec{x}) \right) \Phi_{i_2}(\vec{x}) \right] 
 } & & 
 \nonumber \\
 & = &
 \int d^3y
 \left[ 
   \frac{\delta \Phi_{i_1}(\vec{x})}{\delta \tilde{\Phi}_{i_2}(\vec{y})} \omega_y \tilde{\Phi}_{i_2}(\vec{y})
   + \frac{\delta \Phi_{i_1}(\vec{x})}{\delta \tilde{W}^j_{\bot}(\vec{y})} \omega_y \tilde{W}^j_{\bot}(\vec{y})
   + \frac{\delta \Phi_{i_1}(\vec{x})}{\delta \tilde{B}_{\bot}(\vec{y})} \omega_y \tilde{B}_{\bot}(\vec{y})
 \right].
\hspace*{40mm}
\eq
To solve these equations it is simplest to combine the $U(1)$-field $B_\mu$ and the $SU(2)$-field $W^j_\mu$ into
a $U(2)$-field $V^a_\mu=(B_\mu,W^j_\mu)$, where the index $a$ takes values from $0$ to $3$.
If the two couplings $g$ and $g'$ would be equal, we would have a perfect $U(2)$-gauge theory coupled to a scalar field.
The fact that the two couplings are not equal leads only to minor complication which we can deal with by adjusting
in the appropriate places the coupling factors.
To this aim we define by
\bq
 n_0\left(a_1,...,a_n\right)
\eq
the number of times a zero occurs in the list $a_1,...,a_n$.
We observe that the gauge fields occur in eq.~(\ref{Lagrangian_transverse2}) in ${\cal L}_{+--}$ and ${\cal L}_{++--}$ 
either in a combination like
\bq
 {\bf W}_{\bot} + {\bf B}_{\bot}
 & = & 
 -i \left( g I^k W^k_{\bot} + g' I^0 B_{\bot} \right)
 =
 - i g \left( \frac{g'}{g} \right)^{n_0(a)} I^a V^a_{\bot}
\eq
or in commutators to which only the $SU(2)$-gauge field give a non-vanishing contribution.
An example is given by the term
\bq
  \frac{4}{g^2} \mbox{Tr} \;
     \left( \partial_{\bot} {\bf W}_{\bot^\ast} \right) \partial_\myminus^{-1} \left[ {\bf W}_{\bot^\ast}, \partial_\myminus {\bf W}_{\bot} \right]
 = 
 4 i g \mbox{Tr} \left( I^a [ I^b, I^c ] \right)
     \left( \frac{g'}{g} \right)^{n_0(a,b,c)}
     \left( \partial_{\bot} V^a_{\bot^\ast} \right) \partial_\myminus^{-1} \left( V^b_{\bot^\ast} \partial_\myminus V^c_{\bot} \right).
 \nonumber
\eq
The inclusion of the factor which adjusts the couplings has no effect here: In all cases where $n_0(a,b,c)$ is non-zero
the accompanying trace is zero. We can summarise these observations in the rule that the $U(2)$-gauge field $V^a_\mu$ is always
accompanied by a factor $(g'/g)^{n_0(a)}$.
In appendix~\ref{sect:solution_coeff_canonical_trafo} we have collected detailed information how the equations of the
canonical transformation are solved.
The solution to the integro-differential equations~(\ref{eq_canonical_trafo}) is given by
\bq
\label{solution_canonical_trafo}
 V_{\bot}^a\left(\vec{x}\right)
 & = &
 \sum\limits_{n=1}^\infty 
 2 \; \mbox{Tr} \left( I^a I^{a_1} ... I^{a_n} \right)
 \left( \frac{g'}{g} \right)^{n_0(a_1,...,a_n)-n_0(a)}
 \int \frac{d^3p_1}{(2\pi)^3} ... \frac{d^3p_n}{(2\pi)^3}
 e^{-i (\vec{p}_1 + ... + \vec{p}_n ) \cdot \vec{x}} 
 \nonumber \\
 & &
 \Upsilon\left(\vec{p}_1,...,\vec{p}_n\right)
 \tilde{V}_{\bot}^{a_1}\left(\vec{p}_1\right) ... \tilde{V}_{\bot}^{a_n}\left(\vec{p}_n\right),
 \\
 \Phi_{i_1}\left(\vec{x}\right)
 & = &
 \sum\limits_{n=1}^\infty
 \left( I^{a_1} ... I^{a_{n-1}} \right)_{i_1 i_2}
 \left( \frac{g'}{g} \right)^{n_0(a_1,...,a_{n-1})}
 \int \frac{d^3p_1}{(2\pi)^3} ... \frac{d^3p_n}{(2\pi)^3}
 e^{-i\left( \vec{p}_1 + ... + \vec{p}_n \right) \cdot \vec{x}}
 \nonumber \\
 & &
 {\cal Z}\left(\vec{p}_1,...,\vec{p}_n\right)
 \tilde{V}^{a_1}_{\bot}\left(\vec{p}_1\right) 
 ...
 \tilde{V}^{a_{n-1}}_{\bot}\left(\vec{p}_{n-1}\right) 
 \tilde{\Phi}_{i_2}\left(\vec{p}_n\right),
 \nonumber 
\eq
\bq
 \Phi_{i_2}^\dagger\left(\vec{x}\right)
 & = &
 \sum\limits_{n=1}^\infty
 \left( I^{a_1} ... I^{a_{n-1}} \right)_{i_1 i_2}
 \left( \frac{g'}{g} \right)^{n_0(a_2,...,a_n)}
 \int \frac{d^3p_1}{(2\pi)^3} ... \frac{d^3p_n}{(2\pi)^3}
 e^{-i\left( \vec{p}_1 + ... + \vec{p}_n \right) \cdot \vec{x}}
 \nonumber \\
 & &
 {\cal X}\left(\vec{p}_1,...,\vec{p}_n\right)
 \tilde{\Phi}_{i_1}^\dagger\left(\vec{p}_1\right)
 \tilde{V}^{a_2}_{\bot}\left(\vec{p}_2\right) 
 ...
 \tilde{V}^{a_{n}}_{\bot}\left(\vec{p}_{n}\right),
 \nonumber \\
 V_{\bot^\ast}^a\left(\vec{x}\right)
 & = & 
 \sum\limits_{n=1}^\infty 
 \sum\limits_{r=1}^n
 2 \; \mbox{Tr} \left( I^a I^{a_1} ... I^{a_n} \right)
 \left( \frac{g'}{g} \right)^{n_0(a_1,...,a_n)-n_0(a)}
 \int \frac{d^3p_1}{(2\pi)^3} ... \frac{d^3p_n}{(2\pi)^3}
 e^{-i (\vec{p}_1 + ... + \vec{p}_n ) \cdot \vec{x}} 
 \nonumber \\
 & &
 \Xi_r\left(\vec{p}_1,...,\vec{p}_n\right)
 \tilde{V}_{\bot}^{a_1}\left(\vec{p}_1\right) 
 ... 
 \tilde{V}_{\bot}^{a_{r-1}}\left(\vec{p}_{r-1}\right) 
 \tilde{V}_{\bot^\ast}^{a_r}\left(\vec{p}_r\right) 
 \tilde{V}_{\bot}^{a_{r+1}}\left(\vec{p}_{r+1}\right) 
 ... 
 \tilde{V}_{\bot}^{a_n}\left(\vec{p}_n\right)
 \nonumber \\
 & &
 +
 \sum\limits_{n=2}^\infty
 \sum\limits_{r=1}^{n-1}
 \left( I^{a_{r+2}} ... I^{a_n} I^a I^{a_1} ... I^{a_{r-1}} \right)_{i_1 i_2}
 \left( \frac{g'}{g} \right)^{n_0(a_1,...,a_{r-1},a_{r+2},...,a_n)-n_0(a)}
 \nonumber \\
 & &
 \int \frac{d^3p_1}{(2\pi)^3} ... \frac{d^3p_n}{(2\pi)^3}
 e^{-i (\vec{p}_1 + ... + \vec{p}_n ) \cdot \vec{x}} 
 \nonumber \\
 & &
 \Omega_r\left( \vec{p_1}, ..., \vec{p}_n \right)
 \tilde{V}_{\bot}^{a_1}\left(\vec{p}_1\right)
 ...
 \tilde{V}_{\bot}^{a_{r-1}}\left(\vec{p}_{r-1}\right)
 \tilde{\Phi}_{i_2}\left(\vec{p}_r\right)
 \tilde{\Phi}_{i_1}^\dagger\left(\vec{p}_{r+1}\right)
 \tilde{V}_{\bot}^{a_{r+2}}\left(\vec{p}_{r+2}\right)
 ...
 \tilde{V}_{\bot}^{a_n}\left(\vec{p}_n\right).
 \nonumber 
\eq
The coefficient functions are given by
\bq
\label{coefficient_functions}
 \Upsilon\left(\vec{p}_1,...,\vec{p}_n\right) 
 & = &
 \frac{\left( \mynosign \sqrt{2} g \right)^{n-1}}{\mylbracket p_1 p_2 \myrbracket ... \mylbracket p_{n-1} p_n \myrbracket} \frac{p_1^\myminus+...+p_n^\myminus}{\sqrt{p_1^\myminus p_n^\myminus}},
 \nonumber \\
 {\cal Z}\left(\vec{p}_1,...,\vec{p}_n\right)
 & = & \frac{p_n^\myminus}{p_1^\myminus+...+p_n^\myminus} \Upsilon\left(\vec{p}_1,...,\vec{p}_n\right),
 \nonumber \\
 {\cal X}\left(\vec{p}_1,...,\vec{p}_n\right)
 & = & \frac{p_1^\myminus}{p_1^\myminus+...+p_n^\myminus} \Upsilon\left(\vec{p}_1,...,\vec{p}_n\right),
 \nonumber \\
 \Xi_r\left(\vec{p}_1,...,\vec{p}_n\right) 
 & = & \left( \frac{p_r^\myminus}{p_1^\myminus+...+p_n^\myminus} \right)^2 \Upsilon\left(\vec{p}_1,...,\vec{p}_n\right),
 \nonumber \\
 \Omega_r\left( \vec{p_1}, ..., \vec{p}_n \right)
 & = &
 - \frac{p_r^\myminus p_{r+1}^\myminus}{\left(p_1^\myminus+...+p_n^\myminus\right)^2}
 \Upsilon\left(\vec{p}_1,...,\vec{p}_n\right).
\eq
We remark that the field $V_{\bot}^a\left(\vec{x}\right)$ is expressed in terms of the fields
$\tilde{V}_{\bot}^{a}\left(\vec{p}\right)$ alone, while the field
$V_{\bot^\ast}^a\left(\vec{x}\right)$ involves not only 
$\tilde{V}_{\bot^\ast}^{a}\left(\vec{p}\right)$ and
$\tilde{V}_{\bot}^{a}\left(\vec{p}\right)$, but also the scalar fields
$\tilde{\Phi}_{i_1}^\dagger\left(\vec{p}\right)$ and
$\tilde{\Phi}_{i_2}\left(\vec{p}\right)$.
In all cases the new fields agree with the old fields to leading order in $g$ and $g'$:
\bq
\label{simple_equivalence}
 V_{\bot}^a\left(\vec{x}\right)
 = 
 \tilde{V}_{\bot}^{a}\left(\vec{x}\right) + {\cal O}\left(g,g'\right),
 & &
 V_{\bot^\ast}^a\left(\vec{x}\right)
 =  
 \tilde{V}_{\bot^\ast}^{a}\left(\vec{x}\right) + {\cal O}\left(g,g'\right),
 \nonumber \\
 \Phi_{i_2}\left(\vec{x}\right)
 = 
 \tilde{\Phi}_{i_2}\left(\vec{x}\right) + {\cal O}\left(g,g'\right),
 & &
 \Phi_{i_1}^\dagger\left(\vec{x}\right)
 = 
 \tilde{\Phi}_{i_1}^\dagger\left(\vec{x}\right) + {\cal O}\left(g,g'\right).
\eq

\subsection{Step 5: Assembling the pieces}

We are now in a position to put all the pieces together.
Inserting the solutions~(\ref{solution_canonical_trafo}) 
of the canonical transformation into the Lagrange density~(\ref{Lagrangian_transverse3})
one finds that the Lagrange density can be written in the following form:
\bq
\label{Lagrangian_broken_MHV}
 {\cal L}_{\mathrm{EW}} & = &
 {\cal L}_{\mathrm{kin}} 
 + {\cal L}^{(n)} + {\cal L}^{(n)}_{\bar{\Phi}\Phi} + {\cal L}^{(n)}_{\bar{\Phi}\Phi\bar{\Phi}\Phi}
 + {\cal L}^{(n)}_\mu
 + {\cal L}^{(n)}_{\bar{\Phi}_0\Phi} + {\cal L}^{(n)}_{\bar{\Phi}\Phi_0} + {\cal L}^{(n)}_{\bar{\Phi}_0\Phi_0} 
 \nonumber \\
 & &
 + {\cal L}^{(n)}_{\bar{\Phi}_0\Phi\bar{\Phi}\Phi} 
 + {\cal L}^{(n)}_{\bar{\Phi}\Phi\bar{\Phi}\Phi_0} 
 + {\cal L}^{(n)}_{\bar{\Phi}_0\Phi\bar{\Phi}\Phi_0} 
 + {\cal L}^{(n)}_{\bar{\Phi}_0\Phi\bar{\Phi}_0\Phi} 
 + {\cal L}^{(n)}_{\bar{\Phi}\Phi_0\bar{\Phi}\Phi_0} 
 + {\cal L}^{(n)}_{\bar{\Phi}\Phi\bar{\Phi}_0\Phi_0}.
\eq
The first term ${\cal L}_{\mathrm{kin}}$ is rather simple and contains the kinetic terms:
\bq
\label{Lagrangian_kinetic}
 {\cal L}_{\mathrm{kin}}
 & = &
 - \tilde{V}^a_{\bot^\ast}(x) \Box \tilde{V}^a_{\bot}(x)
 - \tilde{\Phi}^\dagger(x) \Box \tilde{\Phi}(x).
\eq
All other terms contain each an ascending tower of interaction vertices.
Each interaction vertex is most conveniently expressed with the help of the Fourier transforms.
The series of interaction vertices contained in ${\cal L}^{(n)}$ involves only gauge fields.
One finds
\bq
 {\cal L}^{(n)}
 & = & 
 \frac{1}{2}
   \sum\limits_{n=3}^\infty \sum\limits_{j=2}^n
   \int\limits_{(1,..,n)} dP(x) \;
   \alpha_{j}\left(p_1,...,p_n\right)
 \nonumber \\
 & &
  2 \mbox{Tr}\left( 
   {\bf \tilde{V}}_{\bot^\ast}(p_1) {\bf \tilde{V}}_{\bot}(p_2) ... {\bf \tilde{V}}_{\bot}(p_{j-1}) 
   {\bf \tilde{V}}_{\bot^\ast}(p_j) 
   {\bf \tilde{V}}_{\bot}(p_{j+1}) ... {\bf \tilde{V}}_{\bot}(p_{n}) \right).
\eq
The vertex function $\alpha_{j}\left(p_1,...,p_n\right)$ is given 
\bq
 \alpha_{j}\left(p_1,...,p_n\right) 
 & = &
 - \frac{1}{g^2}
 \left( \mynosign i \sqrt{2} \right)^{n-2}
 \frac{\mylbracket p_1 p_j \myrbracket^4}{\mylbracket p_1 p_2 \myrbracket \mylbracket p_2 p_3 \myrbracket ... \mylbracket p_{n-1} p_n \myrbracket \mylbracket p_n p_1 \myrbracket}
\eq
and corresponds exactly to the MHV formula.
Each vertex contains two fields $V_{\bot^\ast}$ with indices $1$ and $j$ and an arbitrary number of fields $V_{\bot}$.
Since the trace is cyclic, we have
\bq
 \mbox{Tr}\left( I^{a_1} ... I^{a_{j-1}} I^{a_j} ... I^{a_n} \right) & = &
  \mbox{Tr}\left( I^{a_j} ... I^{a_n} I^{a_1} ... I^{a_{j-1}} \right).
\eq
The factor $1/2$ takes into account that we are summing twice over identical traces.
The third term ${\cal L}^{(n)}_{\bar{\Phi}\Phi}$ contains two scalar fields and an arbitrary number of gauge fields.
This term reads
\bq
{\cal L}^{(n)}_{\bar{\Phi}\Phi}
 & = &
   \sum\limits_{n=3}^\infty \sum\limits_{j=2}^{n-1}
   \int\limits_{(1,..,n)} dP(x) \;
   \beta_{j}\left(p_1,...,p_n\right)
 \nonumber \\
 & &
   \tilde{\Phi}^\dagger(p_1) {\bf \tilde{V}}_{\bot}(p_2) ... {\bf \tilde{V}}_{\bot}(p_{j-1}) 
   {\bf \tilde{V}}_{\bot^\ast}(p_j) 
   {\bf \tilde{V}}_{\bot}(p_{j+1}) ... {\bf \tilde{V}}_{\bot}(p_{n-1}) \tilde{\Phi}(p_n).
\eq
The coefficient function $\beta_{j}\left(p_1,...,p_n\right)$ is given by
\bq
 \beta_{j}\left(p_1,...,p_n\right) 
 & = &
 -
 \left( \mynosign i \sqrt{2} \right)^{n-2}
 \frac{\mylbracket p_1 p_j \myrbracket^2 \mylbracket p_j p_n \myrbracket^2}{\mylbracket p_1 p_2 \myrbracket \mylbracket p_2 p_3 \myrbracket ... \mylbracket p_{n-1} p_n \myrbracket \mylbracket p_n p_1 \myrbracket}.
\eq
Each vertex contains exactly one field $\tilde{\Phi}^\dagger$ and one field $\tilde{V}_{\bot^\ast}$.
These fields are counted as ``-''. The remaining fields of the vertex are one field $\tilde{\Phi}$ and an
arbitrary number of fields $V_{\bot}$, which are all counted as ``+''.
The vertices correspond therefore to MHV vertices.

The term ${\cal L}^{(n)}_{\bar{\Phi}\Phi\bar{\Phi}\Phi}$ contains four scalar fields plus an arbitrary number of gauge fields.
It is given by
\bq
 {\cal L}^{(n)}_{\bar{\Phi}\Phi\bar{\Phi}\Phi}
 & = &
   \frac{1}{2}
   \sum\limits_{n=4}^\infty \sum\limits_{j=3}^{n-1}
   \int\limits_{(1,..,n)} dP(x) \;
   \left( \gamma_{j}\left(p_1,...,p_n\right) + \delta_{j}\left(p_1,...,p_n\right) + \lambda_{j}\left(p_1,...,p_n\right) \right)
 \nonumber \\
 & &
   \tilde{\Phi}^\dagger(p_1) 
   {\bf \tilde{V}}_{\bot}(p_2) ... {\bf \tilde{V}}_{\bot}(p_{j-2}) 
   \tilde{\Phi}(p_{j-1}) \tilde{\Phi}^\dagger(p_j) 
   {\bf \tilde{V}}_{\bot}(p_{j+1}) ... {\bf \tilde{V}}_{\bot}(p_{n-1}) \tilde{\Phi}(p_n).
\eq
The vertices are again MHV vertices, the two $\tilde{\Phi}^\dagger$-fields are counted as ``-'', all other fields are of
the type ``+''.
We have written explicitly a factor $1/2$ in front, since we sum twice over identical strings of generators of the gauge
group.
We have here three vertex functions
$\gamma_{j}\left(p_1,...,p_n\right)$, $\delta_{j}\left(p_1,...,p_n\right)$ and $\lambda_{j}\left(p_1,...,p_n\right)$.
The explicit form of these functions is given by
\bq
\lefteqn{
 \gamma_{j}\left(p_1,...,p_n\right) 
 = 
 -
 \frac{g^2}{4}
 \left( \mynosign i \sqrt{2} \right)^{n-2}
 \frac{\mylbracket p_1 p_{j-1} \myrbracket^2 \mylbracket p_j p_n \myrbracket^2}
      {\mylbracket p_1 p_2 \myrbracket \mylbracket p_2 p_3 \myrbracket ... \mylbracket p_{n-1} p_n \myrbracket \mylbracket p_n p_1 \myrbracket}
 \left( 1 + \frac{\mylbracket p_1 p_{j} \myrbracket \mylbracket p_{j-1} p_n \myrbracket}{\mylbracket p_1 p_{j-1} \myrbracket \mylbracket p_j p_n \myrbracket} \right),
} & &
 \nonumber \\
\lefteqn{
 \delta_{j}\left(p_1,...,p_n\right) 
 = 
 \frac{1}{4} \left( g'^2 - g^2 \right)
 \left( \mynosign i \sqrt{2} \right)^{n-4}
 \frac{\mylbracket p_{j-1} p_j \myrbracket \mylbracket p_n p_1 \myrbracket}
      {\mylbracket p_1 p_2 \myrbracket \mylbracket p_2 p_3 \myrbracket ... \mylbracket p_{n-1} p_n \myrbracket \mylbracket p_n p_1 \myrbracket}
 \frac{1}{\sqrt{p_{1..j-1}^\myminus p_{j..n}^\myminus}}
} & &
 \nonumber \\
 & &
 \left( \sqrt{p_1^\myminus} \mylbracket p_{1..j-1} p_{j-1} \myrbracket - \sqrt{p_{j-1}^\myminus} \mylbracket p_1 p_{1..j-1} \myrbracket \right)
 \left( \sqrt{p_j^\myminus} \mylbracket p_{j..n} p_{n} \myrbracket - \sqrt{p_{n}^\myminus} \mylbracket p_j p_{j..n} \myrbracket \right),
 \nonumber \\
\lefteqn{
 \lambda_{j}\left(p_1,...,p_n\right) 
 = 
 - \frac{1}{2} \lambda \left( \mynosign i \sqrt{2} \right)^{n-4}
 \frac{\mylbracket p_1 p_{j-1} \myrbracket \mylbracket p_1 p_{n} \myrbracket \mylbracket p_j p_{j-1} \myrbracket \mylbracket p_j p_{n} \myrbracket}
      {\mylbracket p_1 p_2 \myrbracket \mylbracket p_2 p_3 \myrbracket ... \mylbracket p_{n-1} p_n \myrbracket \mylbracket p_n p_1 \myrbracket}.
} & &
\eq
Here we used the short-hand notation
\bq
 p_{i..j} & = & p_i + p_{i+1} + ... + p_j.
\eq
$\gamma_{j}\left(p_1,...,p_n\right)$ arises from the minimal coupling of the scalar field to a $U(2)$ gauge theory.
The vertex function $\delta_{j}\left(p_1,...,p_n\right)$ is proportional to $(g'^2-g^2)$ and arises from the last term
of ${\cal L}_{++--}$ in eq.~(\ref{Lagrangian_transverse2}).
Finally, $\lambda_{j}\left(p_1,...,p_n\right)$ results from the $(\Phi^\dagger \Phi)^2$-term in the Higgs potential.

The piece ${\cal L}^{(n)}_\mu$ of eq.~(\ref{Lagrangian_broken_MHV}) is obtained from the quadratic term
in the Higgs potential. It reads
\bq
 {\cal L}^{(n)}_\mu
 & = &
   \sum\limits_{n=2}^\infty 
   \int\limits_{(1,..,n)} dP(x)
   \mu\left(p_1,...,p_n\right) 
   \tilde{\Phi}^\dagger(p_1) {\bf \tilde{V}}_{\bot}(p_2) ... 
   {\bf \tilde{V}}_{\bot}(p_{n-1}) \tilde{\Phi}(p_n).
\eq
The coefficient function is given by
\bq
 \mu\left(p_1,...,p_n\right) 
 & = &
 - \mu^2 \left( \mynosign i \sqrt{2} \right)^{n-2}
 \frac{\mylbracket p_1 p_n \myrbracket^2}
      {\mylbracket p_1 p_2 \myrbracket \mylbracket p_2 p_3 \myrbracket ... \mylbracket p_{n-1} p_n \myrbracket \mylbracket p_n p_1 \myrbracket}.
\eq
Note that the $n=2$ contribution is the standard mass term for the scalar field:
\bq
 {\cal L}^{(2)}_\mu
 & = &
 \mu^2 \tilde{\Phi}^\dagger(x) \tilde{\Phi}(x).
\eq
Up to now all expressions would equally apply to an unbroken gauge theory coupled to a scalar field with a
quartic self-interaction. The theory is unbroken if $m^2 = - \mu^2 > 0$.
The remaining pieces in the Lagrangian of eq.~(\ref{Lagrangian_broken_MHV})
are all related to the spontaneously symmetry breaking and proportional to $v$
or $v^2$.
The terms
${\cal L}^{(n)}_{\bar{\Phi}_0\Phi}$, ${\cal L}^{(n)}_{\bar{\Phi}\Phi_0}$ and ${\cal L}^{(n)}_{\bar{\Phi}_0\Phi_0}$
read
\bq
{\cal L}^{(n)}_{\bar{\Phi}_0\Phi}
 & = &
   \sum\limits_{n=4}^\infty \sum\limits_{j=2}^{n-1}
   \int\limits_{(2,..,n)} dP(x) \;
   \beta^{(1)}_{j}\left(p_2,...,p_n\right)
\nonumber \\
 &&
   \Phi_{0}^\dagger {\bf \tilde{V}}_{\bot}(p_2) ... {\bf \tilde{V}}_{\bot}(p_{j-1}) {\bf \tilde{V}}_{\bot^\ast}(p_j) 
   {\bf \tilde{V}}_{\bot}(p_{j+1}) ... {\bf \tilde{V}}_{\bot}(p_{n-1}) \tilde{\Phi}(p_n),
 \nonumber \\
{\cal L}^{(n)}_{\bar{\Phi}\Phi_0}
 & = &
   \sum\limits_{n=4}^\infty \sum\limits_{j=2}^{n-1}
   \int\limits_{(1,..,n-1)} dP(x) \;
   \beta^{(n)}_{j}\left(p_1,...,p_{n-1}\right)
\nonumber \\
 &&
   \tilde{\Phi}^\dagger(p_1) {\bf \tilde{V}}_{\bot}(p_2) ... {\bf \tilde{V}}_{\bot}(p_{j-1}) {\bf \tilde{V}}_{\bot^\ast}(p_j) 
   {\bf \tilde{V}}_{\bot}(p_{j+1}) ... {\bf \tilde{V}}_{\bot}(p_{n-1}) \Phi_{0},
 \nonumber \\
{\cal L}^{(n)}_{\bar{\Phi}_0\Phi_0}
 & = &
   \sum\limits_{n=4}^\infty \sum\limits_{j=2}^{n-1}
   \int\limits_{(2,..,n-1)} dP(x) \;
   \beta^{(1,n)}_{j}\left(p_2,...,p_{n-1}\right)
\nonumber \\
 &&
   \Phi_{0}^\dagger {\bf \tilde{V}}_{\bot}(p_2) ... {\bf \tilde{V}}_{\bot}(p_{j-1}) {\bf \tilde{V}}_{\bot^\ast}(p_j) 
   {\bf \tilde{V}}_{\bot}(p_{j+1}) ... {\bf \tilde{V}}_{\bot}(p_{n-1}) \Phi_{0}.
\eq
The coefficient functions are given by
\bq
 \beta^{(1)}_{j}\left(p_2,...,p_n\right) 
 & = &
 \left( \mynosign i \sqrt{2} \right)^{n-2}
 \frac{p_j^\myminus}{\sqrt{p_2^\myminus p_n^\myminus}}
 \frac{\mylbracket p_j p_n \myrbracket^2}
      {\mylbracket p_2 p_3 \myrbracket \mylbracket p_3 p_4 \myrbracket ... \mylbracket p_{n-1} p_n \myrbracket},
 \nonumber \\
 \beta^{(n)}_{j}\left(p_1,...,p_{n-1}\right) 
 & = &
 \left( \mynosign i \sqrt{2} \right)^{n-2}
 \frac{p_j^\myminus}{\sqrt{p_1^\myminus p_{n-1}^\myminus}}
 \frac{\mylbracket p_1 p_j \myrbracket^2}
      {\mylbracket p_1 p_2 \myrbracket \mylbracket p_2 p_3 \myrbracket ... \mylbracket p_{n-2} p_{n-1} \myrbracket},
 \nonumber \\
 \beta^{(1,n)}_{j}\left(p_2,...,p_{n-1}\right) 
 & = &
 \frac{1}{2}
 \left( \mynosign i \sqrt{2} \right)^{n-2}
 \frac{\left(p_j^\myminus\right)^2}{p_2^\myminus p_{n-1}^\myminus}
 \frac{\mylbracket p_2 p_{n-1} \myrbracket}
      {\mylbracket p_2 p_3 \myrbracket \mylbracket p_3 p_4 \myrbracket ... \mylbracket p_{n-2} p_{n-1} \myrbracket}.
\eq
The term ${\cal L}^{(4)}_{\bar{\Phi}_0\Phi_0}$ provides the masses for the electro-weak gauge bosons.
Using momentum conservation the corresponding coefficient functions simplify to
\bq
 \beta^{(1,4)}_{2}\left(p_2,p_3\right)
 =
 \beta^{(1,4)}_{3}\left(p_2,p_3\right)
 = 1.
\eq
The remaining terms in the second line of eq.~(\ref{Lagrangian_broken_MHV}) read
\bq
 {\cal L}^{(n)}_{\bar{\Phi}_0\Phi\bar{\Phi}\Phi}
 & = &
   \sum\limits_{n=4}^\infty \sum\limits_{j=3}^{n-1}
   \int\limits_{(2,..,n)} dP(x) \;
   \left( \gamma^{(1)}_{j}\left(p_2,...,p_n\right) + \delta^{(1)}_{j}\left(p_2,...,p_n\right) + \lambda^{(1)}_{j}\left(p_2,...,p_n\right) \right)
 \nonumber \\
 & &
   \Phi_{0}^\dagger 
   {\bf \tilde{V}}_{\bot}(p_2) ... {\bf \tilde{V}}_{\bot}(p_{j-2}) 
   \tilde{\Phi}(p_{j-1}) \tilde{\Phi}^\dagger(p_j) 
   {\bf \tilde{V}}_{\bot}(p_{j+1}) ... {\bf \tilde{V}}_{\bot}(p_{n-1}) \tilde{\Phi}(p_n),
 \nonumber 
\eq
\bq
 {\cal L}^{(n)}_{\bar{\Phi}\Phi\bar{\Phi}\Phi_0}
 & = &
   \sum\limits_{n=4}^\infty \sum\limits_{j=3}^{n-1}
   \int\limits_{(1,..,n-1)} dP(x) \;
   \left( \gamma^{(n)}_{j}\left(p_1,...,p_{n-1}\right) + \delta^{(n)}_{j}\left(p_1,...,p_{n-1}\right) 
 \right. \nonumber \\
 & & \left.
          + \lambda^{(n)}_{j}\left(p_1,...,p_{n-1}\right) \right)
 \nonumber \\
 & &
   \tilde{\Phi}^\dagger(p_1) 
   {\bf \tilde{V}}_{\bot}(p_2) ... {\bf \tilde{V}}_{\bot}(p_{j-2}) 
   \tilde{\Phi}(p_{j-1}) \tilde{\Phi}^\dagger(p_j) 
   {\bf \tilde{V}}_{\bot}(p_{j+1}) ... {\bf \tilde{V}}_{\bot}(p_{n-1}) \Phi_{0},
 \nonumber 
\eq
\bq
 {\cal L}^{(n)}_{\bar{\Phi}_0\Phi\bar{\Phi}\Phi_0}
 & = &
   \sum\limits_{n=4}^\infty \sum\limits_{j=3}^{n-1}
   \int\limits_{(2,..,n-1)} dP(x) \;
   \left( \gamma^{(1,n)}_{j}\left(p_2,...,p_{n-1}\right) + \delta^{(1,n)}_{j}\left(p_2,...,p_{n-1}\right) 
 \right. \nonumber \\
 & & \left. + \lambda^{(1,n)}_{j}\left(p_2,...,p_{n-1}\right) \right)
 \nonumber \\
 & &
   \Phi_{0}^\dagger 
   {\bf \tilde{V}}_{\bot}(p_2) ... {\bf \tilde{V}}_{\bot}(p_{j-2}) 
   \tilde{\Phi}(p_{j-1}) \tilde{\Phi}^\dagger(p_j) 
   {\bf \tilde{V}}_{\bot}(p_{j+1}) ... {\bf \tilde{V}}_{\bot}(p_{n-1}) \Phi_{0},
 \nonumber \\
 {\cal L}^{(n)}_{\bar{\Phi}_0\Phi\bar{\Phi}_0\Phi}
 & = &
   \frac{1}{2}
   \sum\limits_{n=4}^\infty \sum\limits_{j=3}^{n-1}
   \int\limits_{(2,..,j-1,j+1,...,n)} dP(x) \;
   \left( \gamma^{(1,j)}_{j}\left(p_2,...,p_{j-1},p_{j+1},...,p_n\right) 
 \right. \nonumber \\
 & & \left.
          + \delta^{(1,j)}_{j}\left(p_2,...,p_{j-1},p_{j+1},...,p_n\right) + \lambda^{(1,j)}_{j}\left(p_2,...,p_{j-1},p_{j+1},...,p_n\right) \right)
 \nonumber \\
 & &
   \Phi_{0}^\dagger 
   {\bf \tilde{V}}_{\bot}(p_2) ... {\bf \tilde{V}}_{\bot}(p_{j-2}) 
   \tilde{\Phi}(p_{j-1}) \Phi_{0} 
   {\bf \tilde{V}}_{\bot}(p_{j+1}) ... {\bf \tilde{V}}_{\bot}(p_{n-1}) \tilde{\Phi}(p_n),
 \nonumber \\
 {\cal L}^{(n)}_{\bar{\Phi}\Phi_0\bar{\Phi}\Phi_0}
 & = &
   \frac{1}{2}
   \sum\limits_{n=4}^\infty \sum\limits_{j=3}^{n-1}
   \int\limits_{(1,..,j-2,j,...,n-1)} dP(x) \;
   \left( \gamma^{(j-1,n)}_{j}\left(p_1,...,p_{j-2},p_j,...,p_{n-1}\right) 
 \right. \nonumber \\
 & & \left. 
          + \delta^{(j-1,n)}_{j}\left(p_1,...,p_{j-2},p_j,...,p_{n-1}\right) + \lambda^{(j-1,n)}_{j}\left(p_1,...,p_{j-2},p_j,...,p_{n-1}\right) \right)
 \nonumber \\
 & &
   \tilde{\Phi}^\dagger(p_1) 
   {\bf \tilde{V}}_{\bot}(p_2) ... {\bf \tilde{V}}_{\bot}(p_{j-2}) 
   \Phi_{0} \tilde{\Phi}^\dagger(p_j) 
   {\bf \tilde{V}}_{\bot}(p_{j+1}) ... {\bf \tilde{V}}_{\bot}(p_{n-1}) \Phi_{0},
 \nonumber \\
 {\cal L}^{(n)}_{\bar{\Phi}\Phi\bar{\Phi}_0\Phi_0}
 & = &
   \sum\limits_{n=4}^\infty 
   \int\limits_{(1,..,n-2)} dP(x) \;
   \left( \gamma^{(n-1,n)}_{n-1}\left(p_1,...,p_{n-2}\right) 
          + \lambda^{(n-1,n)}_{n-1}\left(p_1,...,p_{n-2}\right) \right)
 \nonumber \\
 & &
   \tilde{\Phi}^\dagger(p_1) 
   {\bf \tilde{V}}_{\bot}(p_2) ... {\bf \tilde{V}}_{\bot}(p_{n-3}) 
   \tilde{\Phi}(p_{n-2}) \Phi_{0}^\dagger \Phi_{0}.
\eq
The vertex functions are given by
\bq
\lefteqn{
 \gamma^{(1)}_{j}\left(p_2,...,p_n\right) 
 = } & &
 \nonumber \\
 & &
 \frac{g^2}{4}
 \left( \mynosign i \sqrt{2} \right)^{n-2}
 \frac{p^\myminus_{j-1}}{\sqrt{p^\myminus_2 p^\myminus_n}}
 \frac{\mylbracket p_j p_n \myrbracket^2}
      {\mylbracket p_2 p_3 \myrbracket \mylbracket p_3 p_4 \myrbracket ... \mylbracket p_{n-1} p_n \myrbracket}
 \left( 1 + \sqrt{\frac{p^\myminus_j}{p^\myminus_{j-1}}} \frac{\mylbracket p_{j-1} p_n \myrbracket}{\mylbracket p_j p_n \myrbracket} \right),
 \nonumber \\
\lefteqn{
 \gamma^{(n)}_{j}\left(p_1,...,p_{n-1}\right) 
 = } & & 
 \nonumber \\
 & &
 \frac{g^2}{4}
 \left( \mynosign i \sqrt{2} \right)^{n-2}
 \frac{p^\myminus_j}{\sqrt{p^\myminus_1 p^\myminus_{n-1}}}
 \frac{\mylbracket p_1 p_{j-1} \myrbracket^2}
      {\mylbracket p_1 p_2 \myrbracket \mylbracket p_2 p_3 \myrbracket ... \mylbracket p_{n-2} p_{n-1} \myrbracket}
 \left( 1 + \sqrt{\frac{p^\myminus_{j-1}}{p^\myminus_j}} \frac{\mylbracket p_1 p_{j} \myrbracket}{\mylbracket p_1 p_{j-1} \myrbracket} \right),
 \nonumber 
\eq
\bq
\lefteqn{
 \gamma^{(1,n)}_{j}\left(p_2,...,p_{n-1}\right) 
 = } & &
 \nonumber \\
 & &
 -\frac{g^2}{4}
 \left( \mynosign i \sqrt{2} \right)^{n-2}
 \sqrt{\frac{p^\myminus_{j-1} p^\myminus_{j}}{p^\myminus_{2} p^\myminus_{n-1}}}
 \frac{1}{\mylbracket p_2 p_3 \myrbracket \mylbracket p_3 p_4 \myrbracket ... \mylbracket p_{n-2} p_{n-1} \myrbracket}
 \left( \mylbracket p_{j-1} p_{j} \myrbracket - \sqrt{\frac{p^\myminus_{j-1} p^\myminus_{j}}{p^\myminus_{2} p^\myminus_{n-1}}} \mylbracket p_{2} p_{n-1} \myrbracket \right),
 \nonumber \\
\lefteqn{
 \gamma^{(1,j)}_{j}\left(p_2,...,p_{j-1},p_{j+1},...,p_n\right) 
 = 
 -\frac{g^2}{4}
 \left( \mynosign i \sqrt{2} \right)^{n-2}
 \sqrt{\frac{p^\myminus_{j-1} p^\myminus_{n}}{p^\myminus_{2} p^\myminus_{j+1}}}
 \frac{\mylbracket p_{j-1} p_{j} \myrbracket \mylbracket p_{j} p_{j+1} \myrbracket}
      {\mylbracket p_2 p_3 \myrbracket \mylbracket p_3 p_4 \myrbracket ... \mylbracket p_{n-1} p_{n} \myrbracket},
} & &
 \nonumber \\
\lefteqn{
 \gamma^{(j-1,n)}_{j}\left(p_1,...,p_{j-2},p_j,...,p_{n-1}\right) 
 = 
 -\frac{g^2}{4}
 \left( \mynosign i \sqrt{2} \right)^{n-2}
 \sqrt{\frac{p^\myminus_{1} p^\myminus_{j}}{p^\myminus_{j-2} p^\myminus_{n-1}}}
 \frac{\mylbracket p_{j-2} p_{j-1} \myrbracket \mylbracket p_{j-1} p_{j} \myrbracket}
      {\mylbracket p_1 p_2 \myrbracket \mylbracket p_2 p_3 \myrbracket ... \mylbracket p_{n-2} p_{n-1} \myrbracket},
} & &
 \nonumber \\
\lefteqn{
 \gamma^{(n-1,n)}_{n-1}\left(p_1,...,p_{n-2}\right) 
 = 
 \frac{g^2}{4}
 \left( \mynosign i \sqrt{2} \right)^{n-2}
 \frac{\mylbracket p_1 p_{n-2} \myrbracket}
      {\mylbracket p_1 p_2 \myrbracket \mylbracket p_2 p_3 \myrbracket ... \mylbracket p_{n-3} p_{n-2} \myrbracket}.
} & &
\eq
The coefficient functions of the $\delta$-series are given by
\bq
\lefteqn{
 \delta^{(1)}_{j}\left(p_2,...,p_n\right) 
 = 
 -\frac{1}{4} \left( g'^2 - g^2 \right)
 \left( \mynosign i \sqrt{2} \right)^{n-4}
} & &
 \nonumber \\
 & &
 \sqrt{\frac{p^\myminus_{j-1}}{p^\myminus_2 p_{j..n}^\myminus}}
 \frac{\mylbracket p_{j-1} p_j \myrbracket}
      {\mylbracket p_2 p_3 \myrbracket \mylbracket p_3 p_4 \myrbracket ... \mylbracket p_{n-1} p_n \myrbracket}
 \left( \sqrt{p_j^\myminus} \mylbracket p_{j..n} p_{n} \myrbracket - \sqrt{p_{n}^\myminus} \mylbracket p_j p_{j..n} \myrbracket \right),
 \nonumber \\
\lefteqn{
 \delta^{(n)}_{j}\left(p_1,...,p_{n-1}\right) 
  = 
 \frac{1}{4} \left( g'^2 - g^2 \right)
 \left( \mynosign i \sqrt{2} \right)^{n-4}
} & &
 \nonumber \\
 & &
 \sqrt{\frac{p^\myminus_{j}}{p^\myminus_{1..j-1} p_{n-1}^\myminus}}
 \frac{\mylbracket p_{j-1} p_j \myrbracket}
      {\mylbracket p_1 p_2 \myrbracket \mylbracket p_2 p_3 \myrbracket ... \mylbracket p_{n-2} p_{n-1} \myrbracket}
 \left( \sqrt{p_1^\myminus} \mylbracket p_{1..j-1} p_{j-1} \myrbracket - \sqrt{p_{j-1}^\myminus} \mylbracket p_1 p_{1..j-1} \myrbracket \right),
 \nonumber 
\eq
\bq
\lefteqn{
 \delta^{(1,n)}_{j}\left(p_2,...,p_{n-1}\right) 
 = 
 -\frac{1}{4} \left( g'^2 - g^2 \right)
 \left( \mynosign i \sqrt{2} \right)^{n-4}
 \sqrt{\frac{p^\myminus_{j-1} p^\myminus_{j}}{p^\myminus_{2} p_{n-1}^\myminus}}
 \frac{\mylbracket p_{j-1} p_j \myrbracket}
      {\mylbracket p_2 p_3 \myrbracket \mylbracket p_3 p_4 \myrbracket ... \mylbracket p_{n-2} p_{n-1} \myrbracket},
} & &
 \nonumber \\
\lefteqn{
 \delta^{(1,j)}_{j}\left(p_2,...,p_{j-1},p_{j+1},...,p_n\right) 
 = 
} & &
 \nonumber \\
 & &
 \frac{1}{4} \left( g'^2 - g^2 \right)
 \left( \mynosign i \sqrt{2} \right)^{n-4}
 \sqrt{\frac{p^\myminus_{j-1} p^\myminus_{n}}{p^\myminus_{2} p_{j+1}^\myminus}}
 \frac{\mylbracket p_{j-1} p_j \myrbracket \mylbracket p_{j} p_{j+1} \myrbracket}
      {\mylbracket p_2 p_3 \myrbracket \mylbracket p_3 p_4 \myrbracket ... \mylbracket p_{n-1} p_{n} \myrbracket},
\hspace*{43mm}
 \nonumber \\
\lefteqn{
 \delta^{(j-1,n)}_{j}\left(p_1,...,p_{j-2},p_j,...,p_{n-1}\right) 
  = 
} & &
 \nonumber \\
 & &
 \frac{1}{4} \left( g'^2 - g^2 \right)
 \left( \mynosign i \sqrt{2} \right)^{n-4}
 \sqrt{\frac{p^\myminus_{1} p^\myminus_{j}}{p^\myminus_{j-2} p_{n-1}^\myminus}}
 \frac{\mylbracket p_{j-2} p_{j-1} \myrbracket \mylbracket p_{j-1} p_{j} \myrbracket}
      {\mylbracket p_1 p_2 \myrbracket \mylbracket p_2 p_3 \myrbracket ... \mylbracket p_{n-2} p_{n-1} \myrbracket}.
\eq
Finally, the coefficient functions of the $\lambda$-series are given by
\bq
 \lambda^{(1)}_{j}\left(p_2,...,p_n\right) 
 & = &
 \frac{1}{2} \lambda \left( \mynosign i \sqrt{2} \right)^{n-4}
 \sqrt{\frac{p^\myminus_{j-1}}{p^\myminus_{2}}}
 \frac{\mylbracket p_j p_{j-1} \myrbracket \mylbracket p_j p_{n} \myrbracket}
      {\mylbracket p_2 p_3 \myrbracket \mylbracket p_3 p_4 \myrbracket ... \mylbracket p_{n-1} p_n \myrbracket},
 \nonumber \\
 \lambda^{(n)}_{j}\left(p_1,...,p_{n-1}\right) 
 & = &
 \frac{1}{2} \lambda \left( \mynosign i \sqrt{2} \right)^{n-4}
 \sqrt{\frac{p^\myminus_{j}}{p^\myminus_{n-1}}}
 \frac{\mylbracket p_1 p_{j-1} \myrbracket \mylbracket p_j p_{j-1} \myrbracket}
      {\mylbracket p_1 p_2 \myrbracket \mylbracket p_2 p_3 \myrbracket ... \mylbracket p_{n-2} p_{n-1} \myrbracket},
 \nonumber 
\eq
\bq
 \lambda^{(1,n)}_{j}\left(p_2,...,p_{n-1}\right) 
 & = &
 \frac{1}{2} \lambda \left( \mynosign i \sqrt{2} \right)^{n-4}
 \sqrt{\frac{p^\myminus_{j-1} p^\myminus_{j}}{p^\myminus_{2} p^\myminus_{n-1}}}
 \frac{\mylbracket p_j p_{j-1} \myrbracket}
      {\mylbracket p_2 p_3 \myrbracket \mylbracket p_3 p_4 \myrbracket ... \mylbracket p_{n-2} p_{n-1} \myrbracket},
 \nonumber 
\eq
\bq
 \lambda^{(1,j)}_{j}\left(p_2,...,p_{j-1},p_{j+1},...,p_n\right) 
 & = &
 -\frac{1}{2} \lambda \left( \mynosign i \sqrt{2} \right)^{n-4}
 \sqrt{\frac{p^\myminus_{j-1} p^\myminus_n}{p^\myminus_{2} p^\myminus_{j+1}}}
 \frac{\mylbracket p_{j-1} p_{j} \myrbracket \mylbracket p_j p_{j+1} \myrbracket}
      {\mylbracket p_2 p_3 \myrbracket \mylbracket p_3 p_4 \myrbracket ... \mylbracket p_{n-1} p_n \myrbracket},
 \nonumber \\
 \lambda^{(j-1,n)}_{j}\left(p_1,...,p_{j-2},p_j,...,p_{n-1}\right) 
 & = &
 -\frac{1}{2} \lambda \left( \mynosign i \sqrt{2} \right)^{n-4}
 \sqrt{\frac{p^\myminus_{1} p^\myminus_j}{p^\myminus_{j-2} p^\myminus_{n-1}}}
 \nonumber \\
 & &
 \frac{\mylbracket p_{j-2} p_{j-1} \myrbracket \mylbracket p_{j-1} p_{j} \myrbracket}
      {\mylbracket p_1 p_2 \myrbracket \mylbracket p_2 p_3 \myrbracket ... \mylbracket p_{n-2} p_{n-1} \myrbracket},
 \nonumber \\
 \lambda^{(n-1,n)}_{n-1}\left(p_1,...,p_{n-2}\right) 
 & = &
 -\frac{1}{2} \lambda \left( \mynosign i \sqrt{2} \right)^{n-4}
 \frac{\mylbracket p_1 p_{n-2} \myrbracket}
      {\mylbracket p_1 p_2 \myrbracket \mylbracket p_2 p_3 \myrbracket ... \mylbracket p_{n-3} p_{n-2} \myrbracket}.
\label{last_equation}
\eq
This completes the list of all vertex functions for a spontaneously broken gauge theory in the MHV formulation.
We remark that all vertex functions depend only on the light-cone coordinates $p^{\bot^\ast}$ and $p^\myminus$, but not on $p^{\bot}$ and
$p^\myplus$.

Let us look at the terms bilinear in the fields. These are given by
\bq
 {\cal L}_{\mathrm{bilinear}} 
 & = &
 {\cal L}_{\mathrm{kin}} 
 + {\cal L}^{(2)}_\mu
 + {\cal L}^{(4)}_{\bar{\Phi}_0\Phi_0} 
 + {\cal L}^{(4)}_{\bar{\Phi}_0\Phi\bar{\Phi}\Phi_0} 
 + {\cal L}^{(4)}_{\bar{\Phi}_0\Phi\bar{\Phi}_0\Phi} 
 + {\cal L}^{(4)}_{\bar{\Phi}\Phi_0\bar{\Phi}\Phi_0} 
 + {\cal L}^{(4)}_{\bar{\Phi}\Phi\bar{\Phi}_0\Phi_0}.
\eq
We notice that there are no mixing terms between scalars and gauge fields.
This is related to the fact that the term ${\cal L}_{+--}$ in eq.~(\ref{Lagrangian_transverse2}) transforms invariantly under the
transformation given in eq.~(\ref{shift_scalar}).
The terms bilinear in the fields are most conveniently expressed in terms of the mass eigenstates.
We change to a basis of mass eigenstates with a transformation analogously of
eq.~(\ref{change_to_mass_eigenstates}).
In terms of the mass eigenstates we find
\bq
 {\cal L}_{\mathrm{bilinear}} 
 & = &
   \tilde{A}_{\bot^\ast} \left( -\Box \right) \tilde{A}_\bot
 + \tilde{W}^\myminus_{\bot^\ast} \left( -\Box - m_W^2\right) \tilde{W}^\myplus_\bot
 + \tilde{W}^\myplus_{\bot^\ast} \left( -\Box - m_W^2 \right) \tilde{W}^\myminus_\bot
 + \tilde{Z}_{\bot^\ast} \left( -\Box - m_Z^2 \right) \tilde{Z}_\bot
 \nonumber \\
 & &
 + \frac{1}{2} \tilde{\phi}_1 \left( -\Box - m_W^2 \right) \tilde{\phi}_1
 + \frac{1}{2} \tilde{\phi}_2 \left( -\Box - m_W^2 \right) \tilde{\phi}_2
 + \frac{1}{2} \tilde{\chi} \left( -\Box - m_Z^2 \right) \tilde{\chi}
 \nonumber \\
 & &
 + \frac{1}{2} \tilde{H} \left( -\Box - m_H^2 \right) \tilde{H}.
\eq
The masses are given by
\bq
 m_W^2 = \frac{1}{4} v^2 g^2,
 \;\;\;
 m_Z^2 = \frac{1}{4} v^2 \left( g^2 + {g'}^2 \right),
 \;\;\;
 m_H^2 = \frac{1}{2} v^2 \lambda.
\eq
We note that the pseudo-Goldstone fields $\tilde{\phi}_1$, $\tilde{\phi}_2$ and 
$\tilde{\chi}$ have exactly the same mass as the corresponding gauge bosons.
In the MHV approach each gauge field has two transverse degrees of freedom. 
For each gauge field which acquires a mass there is an additional scalar pseudo-Goldstone
field with the same mass, which provides the third degree of freedom.

\section{Conclusions}
\label{sect:conclusions}

In this article we considered a $SU(2) \times U(1)$ gauge theory coupled to a scalar field with a potential
which leads to a spontaneous symmetry breakdown.
Starting from the standard Lagrangian of such a theory we derived an equivalent Lagrangian in the MHV formulation.
Our main results are given in the formulae~(\ref{Lagrangian_broken_MHV}) to (\ref{last_equation}).
These results describe the theory in terms of simple scalar propagators and towers of interaction vertices
with an increasing number of gauge bosons.
The list of the formulae might look at a first sight rather long, but one should keep in mind that these
formulae are valid for an arbitrary number of gauge bosons.
Therefore in processes with a high number of external gauge bosons these formulae lead to a simplification compared
to a standard Feynman diagram approach.


\begin{appendix}

\section{Inverse differential operators}
\label{sect:inverse_differential_operators}

In this appendix we discuss inverse differential operators. For simplicity we do this for functions of one variable.
The generalisation to several variables is straightforward.
Let $f(x)$ be a function with the Fourier representation
\bq
 f(x) & = & 
 \int \frac{dp}{2\pi} e^{-i p x} \tilde{f}(p).
\eq
$\tilde{f}(p)$ denotes here the Fourier transform of $f(x)$.
The ordinary derivative $\partial$ acts on the Fourier representation as
\bq
 \partial f(x)
 & = &
 -i \int \frac{dp}{2\pi} e^{-i p x} p \tilde{f}(p).
\eq
The action of the inverse differential operator $\partial^{-1}$ on $f(x)$ is defined through the Fourier representation
\bq
 \partial^{-1} f(x)
 & = &
 i \int \frac{dp}{2\pi} e^{-i p x} \frac{\tilde{f}(p)}{p}.
\eq
As an example we have for
\bq
 f(x) = \frac{1}{2\pi} e^{-iqx} 
 & \Rightarrow &
 \partial^{-1} f(x) = \frac{i}{q} f(x).
\eq
From this example it follows that there is no product rule for inverse differential operators. 
If $f_1(x) = e^{-iq_1x}/(2\pi)$ and $f_2(x) = e^{-iq_2x}/(2\pi)$
then
\bq
 \partial^{-1} \left[ f_1(x) f_2(x) \right] & = &
 \frac{i}{q_1+q_2} f_1(x) f_2(x),
\eq
but
\bq
 \left[ \partial^{-1} f_1(x) \right] f_2(x) + f_1(x) \left[ \partial^{-1} f_2(x) \right]
 & = &
 \left( \frac{i}{q_1} + \frac{i}{q_2} \right) f_1(x) f_2(x).
\eq
We are interested in function spaces such that the function together with its generalised derivatives (ordinary derivatives
and inverse derivatives) vanishes at infinity.
We define the space ${\cal F}^{m,n}$ as the space of functions $f(x)$ such that
\bq
 \lim\limits_{x \rightarrow \pm \infty} \partial^j f(x) & = & 0, 
 \;\;\;\;\;\;
 m \le j \le n,
 \;\;\;\;\;\;
 m,n \in {\mathbb Z}.
\eq
Obviously we have for $m' \le m$ and $n \le n'$
\bq
 {\cal F}^{m',n'} & \subset & {\cal F}^{m,n}.
\eq
If $f, g \in {\cal F}^{-1,0}$ we may use for the inverse differential operator partial integration without boundary terms:
\bq
\label{partial_integration}
 \int dx \left[ \partial^{-1} f(x) \right] g(x) 
 & = & 
 \int dx \left[ \partial^{-1} f(x) \right] \left[ \partial \partial^{-1} g(x) \right] 
 =
 - \int dx f(x) \left[ \partial^{-1} g(x) \right].
\eq
If $f \in {\cal F}^{-2,0}$ we have
\bq 
\label{total_differential}
 \int dx \; \partial^{-1} f(x) 
 & = &
 \int dx \; \partial \left[ \partial^{-2} f(x) \right] 
 = 0.
\eq


\section{Solution for the coefficients of the canonical transformation}
\label{sect:solution_coeff_canonical_trafo}

In this appendix we give detailed information on how the solution for the canonical transformation is determined.
We have to solve the integro-differential equations~(\ref{eq_canonical_trafo}).
This is most elegantly done by first solving the special case of equal couplings $g'=g$.
The correct couplings are then restored in the final result.
For equal couplings we combine the $SU(2)$- and the $U(1)$-field into a $U(2)$-field, which we denote by
$V^a_\mu$, where the index takes values from $0$ to $3$.
The integro-differential equations which need to be solved read then
\bq
\label{eq_canonical_trafo_U2}
\lefteqn{
 \omega V_{\bot}^a(\vec{x}) - g f^{abc} \left( \zeta V_{\bot}^b(\vec{x}) \right) V_{\bot}^c(\vec{x}) 
 =  
 \int d^3y \frac{\delta V_{\bot}^a(\vec{x})}{\delta \tilde{V}_{\bot}^b(\vec{y})} 
 \omega_y \tilde{V}_{\bot}^b(\vec{y}),
} & & \\
\lefteqn{
 \omega \Phi_{i_1}(\vec{x}) 
 + i g I^a_{i_1 i_2} \left[ \zeta V^a_{\bot}(\vec{x}) \right] \Phi_{i_2}(\vec{x})
 - i g I^a_{i_1 i_2} \zeta \left[ V^a_{\bot}(\vec{x}) \Phi_{i_2}(\vec{x}) \right] 
 = } & & 
 \nonumber \\
 & &
 \hspace*{50mm}
 \int d^3y 
 \left[
   \frac{\delta \Phi_{i_1}(\vec{x})}{\delta \tilde{\Phi}_{i_2}(\vec{y})} \omega_y \tilde{\Phi}_{i_2}(\vec{y})
   + \frac{\delta \Phi_{i_1}(\vec{x})}{\delta \tilde{V}^b_{\bot}(\vec{y})} \omega_y \tilde{V}^b_{\bot}(\vec{y})
 \right].
 \nonumber
\eq
Let us start with the equation for $V_{\bot}^a\left(\vec{x}\right)$.
We make the ansatz
\bq
 V_{\bot}^a\left(\vec{x}\right) & = &
 \tilde{V}_{\bot}^a\left(\vec{x}\right)
 +
 \sum\limits_{n=2}^\infty
 2 \; \mbox{Tr}\left( I^a I^{a_1} ... I^{a_n} \right)
 \int d^3x_1 ... d^3x_n 
 \Upsilon\left( \vec{x}, \vec{x_1}, ..., \vec{x}_n \right)
 \tilde{V}_{\bot}^{a_1}\left(\vec{x}_1\right)
 ...
 \tilde{V}_{\bot}^{a_n}\left(\vec{x}_n\right).
 \nonumber
\eq
We now introduce the Fourier transforms
\bq
 \tilde{V}_{\bot}^{a}(\vec{x})
 & = & 
 \int \frac{d^3p}{(2\pi)^3} e^{-i \vec{p} \cdot \vec{x}}  \tilde{V}_{\bot}^{a}(\vec{p}),
 \\
 \Upsilon(\vec{x}, \vec{x}_1, ..., \vec{x}_n)
 & = &
 \int \frac{d^3p_1}{(2\pi)^3} ... \frac{d^3p_n}{(2\pi)^3}
 e^{-i \vec{p}_1 \cdot (\vec{x}-\vec{x}_1) - ... -i \vec{p}_n \cdot (\vec{x}-\vec{x}_n)}
 \Upsilon(\vec{p}_1, ..., \vec{p}_n).
 \nonumber
\eq
Expressed in terms of the Fourier transforms we obtain:
\bq
 V_{\bot}^a\left(\vec{x}\right)
 & = &
 \sum\limits_{n=1}^\infty 
 2 \; \mbox{Tr} \left( I^a I^{a_1} ... I^{a_n} \right)
 \int \frac{d^3p_1}{(2\pi)^3} ... \frac{d^3p_n}{(2\pi)^3}
 e^{-i (\vec{p}_1 + ... + \vec{p}_n ) \cdot \vec{x}} 
 \nonumber \\
 & &
 \Upsilon\left(\vec{p}_1,...,\vec{p}_n\right)
 \tilde{V}_{\bot}^{a_1}\left(\vec{p}_1\right) ... \tilde{V}_{\bot}^{a_n}\left(\vec{p}_n\right),
\eq
with $\Upsilon(\vec{p})=1$.
The functional derivative is calculated to
\bq
 \frac{\delta V_{\bot}^a(\vec{x})}{\delta \tilde{V}_{\bot}^b(\vec{y})}
 & = &
 \sum\limits_{n=1}^\infty 
 \sum\limits_{r=1}^n
 2 \; \mbox{Tr} \left( I^a I^{a_1} ... I^{a_{r-1}} I^b I^{a_{r+1}} ... I^{a_n} \right)
 \int \frac{d^3p_1}{(2\pi)^3} ... \frac{d^3p_n}{(2\pi)^3}
 e^{-i (\vec{p}_1 + ... + \vec{p}_n ) \cdot \vec{x} + i \vec{p}_r \cdot \vec{y}} 
 \nonumber \\
 & &
 \Upsilon\left(\vec{p}_1,...,\vec{p}_n\right)
 \tilde{V}_{\bot}^{a_1}\left(\vec{p}_1\right) 
 ... 
 \tilde{V}_{\bot}^{a_{r-1}}\left(\vec{p}_{r-1}\right) 
 \tilde{V}_{\bot}^{a_{r+1}}\left(\vec{p}_{r+1}\right) 
 ...
 \tilde{V}_{\bot}^{a_n}\left(\vec{p}_n\right).
\eq
We then plug these expressions into the first equation of~(\ref{eq_canonical_trafo_U2}).
The coefficient of each trace $\mbox{Tr} \left( I^a I^{a_1} ... I^{a_n} \right)$ has to vanish separately.
This leads to the following equation
\bq
\label{eq_coeff_Upsilon}
\lefteqn{
 \left( \omega_{p_1} + ... + \omega_{p_n} \right) \Upsilon\left(\vec{p}_1,...,\vec{p}_n\right)
 = } & &
 \\
 & &
 \omega_{p_1+...+p_n} \Upsilon\left(\vec{p}_1,...,\vec{p}_n\right)
 + 
 i g \sum\limits_{r=1}^{n-1} \left( \zeta_{p_1+...+p_r} - \zeta_{p_{r+1}+...+p_n} \right)
 \Upsilon\left(\vec{p}_1,...,\vec{p}_r\right) \Upsilon\left(\vec{p}_{r+1},...,\vec{p}_n\right).
 \nonumber
\eq
In order to simplify the notation we have set
\bq
 \omega_{p} =  
 e^{i \vec{p} \vec{x} } \omega e^{-i \vec{p} \vec{x} }
 = - i \frac{p^{\bot\ast} p^\bot}{p^\myminus},
 & &
 \zeta_{p} = 
 e^{i \vec{p} \vec{x} } \zeta e^{-i \vec{p} \vec{x} }
 = \frac{p^\bot}{p^\myminus}.
\eq
Eq.~(\ref{eq_coeff_Upsilon})
is a recursion relation for the coefficient functions $\Upsilon\left(\vec{p}_1,...,\vec{p}_n\right)$:
\bq
 \Upsilon\left(\vec{p}_1,...,\vec{p}_n\right) 
 & = &
 i g
 \sum\limits_{r=1}^{n-1} 
 \frac{\zeta_{p_1+...+p_r} - \zeta_{p_{r+1}+...+p_n}}{\omega_{p_1} + ... + \omega_{p_n} - \omega_{p_1+...+p_n}}
 \Upsilon\left(\vec{p}_1,...,\vec{p}_r\right) \Upsilon\left(\vec{p}_{r+1},...,\vec{p}_n\right),
\eq
with $\Upsilon(\vec{p})=1$.
This recursion relation has the solution:
\bq
 \Upsilon\left(\vec{p}_1,...,\vec{p}_n\right) 
 & = &
 \frac{\left( \mynosign \sqrt{2} g \right)^{n-1}}{\mylbracket p_1 p_2 \myrbracket ... \mylbracket p_{n-1} p_n \myrbracket} \frac{p_1^\myminus+...+p_n^\myminus}{\sqrt{p_1^\myminus p_n^\myminus}}.
\eq
The coefficients $\Upsilon$ satisfy a decoupling identity:
\bq
 \Upsilon\left(\vec{p}_a,\vec{p}_1,...,\vec{p}_n\right)
 + 
 \Upsilon\left(\vec{p}_1,\vec{p}_a,\vec{p}_2,...,\vec{p}_n\right)
 +
 ...
 +
 \Upsilon\left(\vec{p}_1,...,\vec{p}_n,\vec{p}_a\right)
 & = & 0.
\eq
For the scalar field we make the ansatz
\bq
\lefteqn{
 \Phi_{i_1}\left(\vec{x}\right)
 =  
 \tilde{\Phi}_{i_1}\left(\vec{x}\right)
} & & \\
 & &
 + 
 \sum\limits_{n=2}^\infty 
 \left( I^{a_1} ... I^{a_{n-1}} \right)_{i_1 i_2}
 \int d^3x_1 ... d^3x_n
 {\cal Z}\left( \vec{x},\vec{x}_1,...,\vec{x}_n \right)
 \tilde{V}^{a_1}_{\bot}\left(\vec{x}_1\right) 
 ...
 \tilde{V}^{a_{n-1}}_{\bot}\left(\vec{x}_{n-1}\right) 
 \tilde{\Phi}_{i_2}\left(\vec{x}_n\right),
 \nonumber
\eq
or equivalently in Fourier space
\bq
 \Phi_{i_1}\left(\vec{x}\right)
 & = &
 \sum\limits_{n=1}^\infty
 \left( I^{a_1} ... I^{a_{n-1}} \right)_{i_1 i_2}
 \int \frac{d^3p_1}{(2\pi)^3} ... \frac{d^3p_n}{(2\pi)^3}
 e^{-i\left( \vec{p}_1 + ... + \vec{p}_n \right) \cdot \vec{x}}
 \nonumber \\
 & &
 {\cal Z}\left(\vec{p}_1,...,\vec{p}_n\right)
 \tilde{V}^{a_1}_{\bot}\left(\vec{p}_1\right) 
 ...
 \tilde{V}^{a_{n-1}}_{\bot}\left(\vec{p}_{n-1}\right) 
 \tilde{\Phi}_{i_2}\left(\vec{p}_n\right),
\eq
with ${\cal Z}(\vec{p})=1$. One then proceeds as in the case for the $U(2)$-gauge field: One first
calculates the functional derivatives and inserts then the ansatz and the functional derivatives into the
second equation of~(\ref{eq_canonical_trafo_U2}). This yields again a recursion relation
for the coefficient ${\cal Z}\left(\vec{p}_1,...,\vec{p}_n\right)$:
\bq
\label{recursion_Zfunc}
 {\cal Z}\left(\vec{p}_1,...,\vec{p}_n\right)
 & = &
 i g
 \sum\limits_{r=1}^{n-1} 
 \frac{\zeta_{p_1+...+p_r} - \zeta_{p_{1}+...+p_n}}{\omega_{p_1} + ... + \omega_{p_n} - \omega_{p_1+...+p_n}}
 \Upsilon\left(\vec{p}_1,...,\vec{p}_r\right) {\cal Z}\left(\vec{p}_{r+1},...,\vec{p}_n\right).
\eq
The recursion relation involves in addition to the coefficient 
functions ${\cal Z}\left(\vec{p}_1,...,\vec{p}_n\right)$ 
also the coefficient functions $\Upsilon\left(\vec{p}_1,...,\vec{p}_n\right)$ associated with the gauge fields.
The latter are already known.
The solution of the recursion relation eq.~(\ref{recursion_Zfunc}) is given by
\bq
 {\cal Z}\left(\vec{p}_1,...,\vec{p}_n\right)
 & = & \frac{p_n^\myminus}{p_1^\myminus+...+p_n^\myminus} \Upsilon\left(\vec{p}_1,...,\vec{p}_n\right).
\eq
In addition we have to express the ``old'' conjugated fields $V^a_{\bot^\ast}\left(\vec{x}\right)$
and $\Phi^\dagger_{i_1}\left(\vec{x}\right)$ in terms of the ``new'' fields.
The relevant equations to be solved are given in eq.~(\ref{new_momenta}). Adapted to the $U(2)$-case 
these equations read
\bq
\label{new_momenta_U2}
 \partial_\myminus \tilde{V}_{\bot^\ast}^a(\vec{x})
 & = & \int d^3y \; \frac{\delta V_{\bot}^b(\vec{y})}{\delta \tilde{V}_{\bot}^a(\vec{x})} 
                 \; \partial_\myminus V_{\bot^\ast}^b(\vec{y})
            + \frac{\delta \Phi_{i_2}(\vec{y})}{\delta \tilde{V}_{\bot}^a(\vec{x})}
                 \; \partial_\myminus \Phi_{i_2}^\dagger(\vec{y}),
 \nonumber \\
 \partial_\myminus \tilde{\Phi}_{i_1}^\dagger(\vec{x}) 
 & = & \int d^3y \frac{\delta \Phi_{i_2}(\vec{y})}{\delta \tilde{\Phi}_{i_1}(\vec{x})}
                 \; \partial_\myminus \Phi_{i_2}^\dagger(\vec{y}).
\eq
It is technically simpler to start with the scalar field $\Phi_{i_2}^\dagger\left(\vec{x}\right)$.
We make the ansatz
\bq
\lefteqn{
 \Phi_{i_2}^\dagger\left(\vec{x}\right)
 =  
 \tilde{\Phi}_{i_2}^\dagger\left(\vec{x}\right)
} & & \\
 & &
 + 
 \sum\limits_{n=2}^\infty 
 \left( I^{a_1} ... I^{a_{n-1}} \right)_{i_1 i_2}
 \int d^3x_1 ... d^3x_n
 {\cal X}\left( \vec{x},\vec{x}_1,...,\vec{x}_n \right)
 \tilde{\Phi}_{i_1}^\dagger\left(\vec{x}_1\right)
 \tilde{V}^{a_2}_{\bot}\left(\vec{x}_2\right) 
 ...
 \tilde{V}^{a_{n}}_{\bot}\left(\vec{x}_{n}\right).
 \nonumber
\eq
Again we transform to Fourier space and we obtain
\bq
 \Phi_{i_2}^\dagger\left(\vec{x}\right)
 & = &
 \sum\limits_{n=1}^\infty
 \left( I^{a_1} ... I^{a_{n-1}} \right)_{i_1 i_2}
 \int \frac{d^3p_1}{(2\pi)^3} ... \frac{d^3p_n}{(2\pi)^3}
 e^{-i\left( \vec{p}_1 + ... + \vec{p}_n \right) \cdot \vec{x}}
 \nonumber \\
 & &
 {\cal X}\left(\vec{p}_1,...,\vec{p}_n\right)
 \tilde{\Phi}_{i_1}^\dagger\left(\vec{p}_1\right)
 \tilde{V}^{a_2}_{\bot}\left(\vec{p}_2\right) 
 ...
 \tilde{V}^{a_{n}}_{\bot}\left(\vec{p}_{n}\right),
\eq
with ${\cal X}\left(\vec{p}\right)=1.$ We insert these expressions into the second equation
of~(\ref{new_momenta_U2}).
This yields a recursion relation for the coefficient function ${\cal X}\left(\vec{p}_1,...,\vec{p}_n\right)$
\bq
 {\cal X}\left(\vec{p}_1,...,\vec{p}_n\right)
 & = &
 - \sum\limits_{r=1}^{n-1}
 \frac{p_1^\myminus + ... + p_r^\myminus}{p_1^\myminus + ... + p_n^\myminus}
 {\cal X}\left(\vec{p}_1,...,\vec{p}_r\right)
 {\cal Z}\left(\vec{p}_{r+1},...,\vec{p}_n,-\sum\limits_{i=1}^{n}\vec{p}_i\right).
\eq
The solution is given by
\bq
 {\cal X}\left(\vec{p}_1,...,\vec{p}_n\right)
 & = & \frac{p_1^\myminus}{p_1^\myminus+...+p_n^\myminus} \Upsilon\left(\vec{p}_1,...,\vec{p}_n\right).
\eq
Finally we consider the field $V_{\bot^\ast}^a\left(\vec{x}\right)$.
We make the ansatz
\bq
 V_{\bot^\ast}^a\left(\vec{x}\right) 
 & = &
 \sum\limits_{n=1}^\infty
 \sum\limits_{r=1}^n
 2 \; \mbox{Tr}\left( I^a I^{a_1} ... I^{a_n} \right)
 \int d^3x_1 ... d^3x_n 
 \\
 & &
 \Xi_r\left( \vec{x}, \vec{x_1}, ..., \vec{x}_n \right)
 \tilde{V}_{\bot}^{a_1}\left(\vec{x}_1\right)
 ...
 \tilde{V}_{\bot}^{a_{r-1}}\left(\vec{x}_{r-1}\right)
 \tilde{V}_{\bot^\ast}^{a_r}\left(\vec{x}_r\right)
 \tilde{V}_{\bot}^{a_{r+1}}\left(\vec{x}_{r+1}\right)
 ...
 \tilde{V}_{\bot}^{a_n}\left(\vec{x}_n\right)
 \nonumber \\
 & &
 +
 \sum\limits_{n=2}^\infty
 \sum\limits_{r=1}^{n-1}
 \left( I^{a_{r+2}} ... I^{a_n} I^a I^{a_1} ... I^{a_{r-1}} \right)_{i_1 i_2}
 \int d^3x_1 ... d^3x_n 
 \nonumber \\
 & &
 \Omega_r\left( \vec{x}, \vec{x_1}, ..., \vec{x}_n \right)
 \tilde{V}_{\bot}^{a_1}\left(\vec{x}_1\right)
 ...
 \tilde{V}_{\bot}^{a_{r-1}}\left(\vec{x}_{r-1}\right)
 \tilde{\Phi}_{i_2}\left(\vec{x}_r\right)
 \tilde{\Phi}_{i_1}^\dagger\left(\vec{x}_{r+1}\right)
 \tilde{V}_{\bot}^{a_{r+2}}\left(\vec{x}_{r+2}\right)
 ...
 \tilde{V}_{\bot}^{a_n}\left(\vec{x}_n\right).
 \nonumber 
\eq
Note that we have to take into account the pure gauge field contribution as well as a contribution
involving the scalar fields.
In Fourier space we have
\bq
 V_{\bot^\ast}^a\left(\vec{x}\right)
 & = & 
 \sum\limits_{n=1}^\infty 
 \sum\limits_{r=1}^n
 2 \; \mbox{Tr} \left( I^a I^{a_1} ... I^{a_n} \right)
 \int \frac{d^3p_1}{(2\pi)^3} ... \frac{d^3p_n}{(2\pi)^3}
 e^{-i (\vec{p}_1 + ... + \vec{p}_n ) \cdot \vec{x}} 
 \\
 & &
 \Xi_r\left(\vec{p}_1,...,\vec{p}_n\right)
 \tilde{V}_{\bot}^{a_1}\left(\vec{p}_1\right) 
 ... 
 \tilde{V}_{\bot}^{a_{r-1}}\left(\vec{p}_{r-1}\right) 
 \tilde{V}_{\bot^\ast}^{a_r}\left(\vec{p}_r\right) 
 \tilde{V}_{\bot}^{a_{r+1}}\left(\vec{p}_{r+1}\right) 
 ... 
 \tilde{V}_{\bot}^{a_n}\left(\vec{p}_n\right)
 \nonumber \\
 & &
 +
 \sum\limits_{n=2}^\infty
 \sum\limits_{r=1}^{n-1}
 \left( I^{a_{r+2}} ... I^{a_n} I^a I^{a_1} ... I^{a_{r-1}} \right)_{i_1 i_2}
 \int \frac{d^3p_1}{(2\pi)^3} ... \frac{d^3p_n}{(2\pi)^3}
 e^{-i (\vec{p}_1 + ... + \vec{p}_n ) \cdot \vec{x}} 
 \nonumber \\
 & &
 \Omega_r\left( \vec{p_1}, ..., \vec{p}_n \right)
 \tilde{V}_{\bot}^{a_1}\left(\vec{p}_1\right)
 ...
 \tilde{V}_{\bot}^{a_{r-1}}\left(\vec{p}_{r-1}\right)
 \tilde{\Phi}_{i_2}\left(\vec{p}_r\right)
 \tilde{\Phi}_{i_1}^\dagger\left(\vec{p}_{r+1}\right)
 \tilde{V}_{\bot}^{a_{r+2}}\left(\vec{p}_{r+2}\right)
 ...
 \tilde{V}_{\bot}^{a_n}\left(\vec{p}_n\right),
 \nonumber 
\eq
with $\Xi_1\left(\vec{p}\right)=1$.
We then insert this ansatz together with the result for
$\Phi^\dagger_{i_1}\left(\vec{x}\right)$ into the first equation of~(\ref{new_momenta_U2}).
Each term in the resulting expression will have either a factor
$\tilde{V}_{\bot^\ast}^a(\vec{p})$
or a factor $\tilde{\Phi}^\dagger_{i_1}\left(\vec{p}\right)$, but not both.
The coefficients of these two parts have to vanish separately.
Let us first focus on $\tilde{V}_{\bot^\ast}^a(\vec{p})$.
For $n>1$ we obtain the equation
\bq
\lefteqn{ 
 0 =
 } & & \\
 & &
 \sum\limits_{i_1=1}^r \sum\limits_{i_2=r+1}^n
 \left( p_{i_1}^\myminus + ... + p_{i_2-1}^\myminus \right)
 \Xi_{r-i_1+1}\left(\vec{p}_{i_1},...,\vec{p}_{i_2-1}\right)
 \Upsilon\left(\vec{p}_{i_2},...\vec{p}_{n-1},-\sum\limits_{i=1}^{n-1}\vec{p}_i,\vec{p}_1,...,\vec{p}_{i_1-1}\right).
 \nonumber 
\eq
This is again a recursion relation for the coefficients $\Xi_r\left(\vec{p}_1,...,\vec{p}_n\right)$. We can rewrite
this equation as
\bq
\lefteqn{
 \Xi_r\left(\vec{p}_1,...,\vec{p}_n\right)
 = } & &
 \nonumber \\
 & & 
 - \sum\limits_{i_1=2}^r 
   \sum\limits_{i_2=r+1}^{n+1}
    \frac{p_{i_1}^\myminus+...+p_{i_2-1}^\myminus}{p_1^\myminus+...+p_n^\myminus}
    \Xi_{r-i_1+1}\left(\vec{p}_{i_1},...,\vec{p}_{i_2-1}\right)
    \Upsilon\left(\vec{p}_{i_2},...\vec{p}_{n},-\sum\limits_{i=1}^{n}\vec{p}_i,\vec{p}_1,...,\vec{p}_{i_1-1}\right)
 \nonumber \\
 & &
 - \sum\limits_{i_2=r+1}^{n}
    \frac{p_{1}^\myminus+...+p_{i_2-1}^\myminus}{p_1^\myminus+...+p_n^\myminus}
    \Xi_{r}\left(\vec{p}_{1},...,\vec{p}_{i_2-1}\right)
    \Upsilon\left(\vec{p}_{i_2},...\vec{p}_{n},-\sum\limits_{i=1}^{n}\vec{p}_i\right).
\eq
The solution is given by
\bq
 \Xi_r\left(\vec{p}_1,...,\vec{p}_n\right) 
 & = & \left( \frac{p_r^\myminus}{p_1^\myminus+...+p_n^\myminus} \right)^2 \Upsilon\left(\vec{p}_1,...,\vec{p}_n\right).
\eq
In a similar way one finds a recursion relation for $\Omega_r\left( \vec{p_1}, ..., \vec{p}_n \right)$
from the terms proportional to $\tilde{\Phi}^\dagger_{i_1}\left(\vec{p}\right)$.
We quote the final result:
\bq
 \Omega_r\left( \vec{p_1}, ..., \vec{p}_n \right)
 & = &
 - \frac{p_r^\myminus p_{r+1}^\myminus}{\left(p_1^\myminus+...+p_n^\myminus\right)^2}
 \Upsilon\left(\vec{p}_1,...,\vec{p}_n\right).
\eq

\end{appendix}

\bibliography{/home/stefanw/notes/biblio}

\begin{thebibliography}{10}

\bibitem{Witten:2003nn}
E.~Witten,
\newblock Commun. Math. Phys. {\bf 252}, 189 (2004), hep-th/0312171.

\bibitem{Cachazo:2004kj}
F.~Cachazo, P.~Svrcek, and E.~Witten,
\newblock JHEP {\bf 09}, 006 (2004), hep-th/0403047.

\bibitem{Parke:1986gb}
S.~J. Parke and T.~R. Taylor,
\newblock Phys. Rev. Lett. {\bf 56}, 2459 (1986).

\bibitem{Britto:2004ap}
R.~Britto, F.~Cachazo, and B.~Feng,
\newblock Nucl. Phys. {\bf B715}, 499 (2005), hep-th/0412308.

\bibitem{Britto:2005fq}
R.~Britto, F.~Cachazo, B.~Feng, and E.~Witten,
\newblock Phys. Rev. Lett. {\bf 94}, 181602 (2005), hep-th/0501052.

\bibitem{Luo:2005rx}
M.-x. Luo and C.-k. Wen,
\newblock JHEP {\bf 03}, 004 (2005), hep-th/0501121.

\bibitem{Luo:2005my}
M.-x. Luo and C.-k. Wen,
\newblock Phys. Rev. {\bf D71}, 091501 (2005), hep-th/0502009.

\bibitem{Britto:2005dg}
R.~Britto, B.~Feng, R.~Roiban, M.~Spradlin, and A.~Volovich,
\newblock Phys. Rev. {\bf D71}, 105017 (2005), hep-th/0503198.

\bibitem{Badger:2005zh}
S.~D. Badger, E.~W.~N. Glover, V.~V. Khoze, and P.~Svrcek,
\newblock JHEP {\bf 07}, 025 (2005), hep-th/0504159.

\bibitem{Forde:2005ue}
D.~Forde and D.~A. Kosower,
\newblock Phys. Rev. {\bf D73}, 065007 (2006), hep-th/0507292.

\bibitem{Quigley:2005cu}
C.~Quigley and M.~Rozali,
\newblock JHEP {\bf 03}, 004 (2006), hep-ph/0510148.

\bibitem{Risager:2005vk}
K.~Risager,
\newblock JHEP {\bf 12}, 003 (2005), hep-th/0508206.

\bibitem{Draggiotis:2005wq}
P.~D. Draggiotis, R.~H.~P. Kleiss, A.~Lazopoulos, and C.~G. Papadopoulos,
\newblock Eur. Phys. J. {\bf C46}, 741 (2006), hep-ph/0511288.

\bibitem{Vaman:2005dt}
D.~Vaman and Y.-P. Yao,
\newblock JHEP {\bf 04}, 030 (2006), hep-th/0512031.

\bibitem{Ozeren:2006ft}
K.~J. Ozeren and W.~J. Stirling,
\newblock Eur. Phys. J. {\bf C48}, 159 (2006), hep-ph/0603071.

\bibitem{Schwinn:2005pi}
C.~Schwinn and S.~Weinzierl,
\newblock JHEP {\bf 05}, 006 (2005), hep-th/0503015.

\bibitem{Schwinn:2006ca}
C.~Schwinn and S.~Weinzierl,
\newblock JHEP {\bf 03}, 030 (2006), hep-th/0602012.

\bibitem{Dinsdale:2006sq}
M.~Dinsdale, M.~Ternick, and S.~Weinzierl,
\newblock JHEP {\bf 03}, 056 (2006), hep-ph/0602204.

\bibitem{Duhr:2006iq}
C.~Duhr, S.~Hoche, and F.~Maltoni,
\newblock JHEP {\bf 08}, 062 (2006), hep-ph/0607057.

\bibitem{Draggiotis:2006er}
P.~Draggiotis {\em et~al.},
\newblock Nucl. Phys. Proc. Suppl. {\bf 160}, 255 (2006), hep-ph/0607034.

\bibitem{deFlorian:2006ek}
D.~de~Florian and J.~Zurita,
\newblock JHEP {\bf 05}, 073 (2006), hep-ph/0605291.

\bibitem{deFlorian:2006vu}
D.~de~Florian and J.~Zurita,
\newblock JHEP {\bf 11}, 080 (2006), hep-ph/0609099.

\bibitem{Rodrigo:2005eu}
G.~Rodrigo,
\newblock JHEP {\bf 09}, 079 (2005), hep-ph/0508138.

\bibitem{Ferrario:2006np}
P.~Ferrario, G.~Rodrigo, and P.~Talavera,
\newblock Phys. Rev. Lett. {\bf 96}, 182001 (2006), hep-th/0602043.

\bibitem{Schwinn:2007ee}
C.~Schwinn and S.~Weinzierl,
\newblock JHEP {\bf 04}, 072 (2007), hep-ph/0703021.

\bibitem{Hall:2007mz}
A.~Hall,
\newblock Phys. Rev. {\bf D77}, 025011 (2008), arXiv:0710.1300.

\bibitem{Schwinn:2008fm}
C.~Schwinn,
\newblock Phys. Rev. {\bf D78}, 085030 (2008), arXiv:0809.1442.

\bibitem{Bidder:2004tx}
S.~J. Bidder, N.~E.~J. Bjerrum-Bohr, L.~J. Dixon, and D.~C. Dunbar,
\newblock Phys. Lett. {\bf B606}, 189 (2005), hep-th/0410296.

\bibitem{Bidder:2004vx}
S.~J. Bidder, N.~E.~J. Bjerrum-Bohr, D.~C. Dunbar, and W.~B. Perkins,
\newblock Phys. Lett. {\bf B608}, 151 (2005), hep-th/0412023.

\bibitem{Bidder:2005ri}
S.~J. Bidder, N.~E.~J. Bjerrum-Bohr, D.~C. Dunbar, and W.~B. Perkins,
\newblock Phys. Lett. {\bf B612}, 75 (2005), hep-th/0502028.

\bibitem{Bedford:2004nh}
J.~Bedford, A.~Brandhuber, B.~J. Spence, and G.~Travaglini,
\newblock Nucl. Phys. {\bf B712}, 59 (2005), hep-th/0412108.

\bibitem{Britto:2005ha}
R.~Britto, E.~Buchbinder, F.~Cachazo, and B.~Feng,
\newblock Phys. Rev. {\bf D72}, 065012 (2005), hep-ph/0503132.

\bibitem{Bern:2005hs}
Z.~Bern, L.~J. Dixon, and D.~A. Kosower,
\newblock Phys. Rev. {\bf D71}, 105013 (2005), hep-th/0501240.

\bibitem{Bern:2005ji}
Z.~Bern, L.~J. Dixon, and D.~A. Kosower,
\newblock Phys. Rev. {\bf D72}, 125003 (2005), hep-ph/0505055.

\bibitem{Bern:2005cq}
Z.~Bern, L.~J. Dixon, and D.~A. Kosower,
\newblock Phys. Rev. {\bf D73}, 065013 (2006), hep-ph/0507005.

\bibitem{Bern:2005hh}
Z.~Bern, N.~E.~J. Bjerrum-Bohr, D.~C. Dunbar, and H.~Ita,
\newblock JHEP {\bf 11}, 027 (2005), hep-ph/0507019.

\bibitem{Forde:2005hh}
D.~Forde and D.~A. Kosower,
\newblock Phys. Rev. {\bf D73}, 061701 (2006), hep-ph/0509358.

\bibitem{Berger:2006ci}
C.~F. Berger, Z.~Bern, L.~J. Dixon, D.~Forde, and D.~A. Kosower,
\newblock Phys. Rev. {\bf D74}, 036009 (2006), hep-ph/0604195.

\bibitem{Berger:2006vq}
C.~F. Berger, Z.~Bern, L.~J. Dixon, D.~Forde, and D.~A. Kosower,
\newblock Phys. Rev. {\bf D75}, 016006 (2007), hep-ph/0607014.

\bibitem{Berger:2006sh}
C.~F. Berger, V.~Del~Duca, and L.~J. Dixon,
\newblock Phys. Rev. {\bf D74}, 094021 (2006), hep-ph/0608180.

\bibitem{Britto:2006sj}
R.~Britto, B.~Feng, and P.~Mastrolia,
\newblock Phys. Rev. {\bf D73}, 105004 (2006), hep-ph/0602178.

\bibitem{Xiao:2006vr}
Z.~Xiao, G.~Yang, and C.-J. Zhu,
\newblock Nucl. Phys. {\bf B758}, 1 (2006), hep-ph/0607015.

\bibitem{Su:2006vs}
X.~Su, Z.~Xiao, G.~Yang, and C.-J. Zhu,
\newblock Nucl. Phys. {\bf B758}, 35 (2006), hep-ph/0607016.

\bibitem{Xiao:2006vt}
Z.~Xiao, G.~Yang, and C.-J. Zhu,
\newblock Nucl. Phys. {\bf B758}, 53 (2006), hep-ph/0607017.

\bibitem{Binoth:2006hk}
T.~Binoth, J.~P. Guillet, and G.~Heinrich,
\newblock JHEP {\bf 02}, 013 (2007), hep-ph/0609054.

\bibitem{Binoth:2007ca}
T.~Binoth, G.~Heinrich, T.~Gehrmann, and P.~Mastrolia,
\newblock Phys. Lett. {\bf B649}, 422 (2007), hep-ph/0703311.

\bibitem{Ossola:2006us}
G.~Ossola, C.~G. Papadopoulos, and R.~Pittau,
\newblock Nucl. Phys. {\bf B763}, 147 (2007), hep-ph/0609007.

\bibitem{Ossola:2007bb}
G.~Ossola, C.~G. Papadopoulos, and R.~Pittau,
\newblock JHEP {\bf 07}, 085 (2007), arXiv:0704.1271.

\bibitem{Anastasiou:2006jv}
C.~Anastasiou, R.~Britto, B.~Feng, Z.~Kunszt, and P.~Mastrolia,
\newblock Phys. Lett. {\bf B645}, 213 (2007), hep-ph/0609191.

\bibitem{Anastasiou:2006gt}
C.~Anastasiou, R.~Britto, B.~Feng, Z.~Kunszt, and P.~Mastrolia,
\newblock JHEP {\bf 03}, 111 (2007), hep-ph/0612277.

\bibitem{Mastrolia:2006ki}
P.~Mastrolia,
\newblock Phys. Lett. {\bf B644}, 272 (2007), hep-th/0611091.

\bibitem{Britto:2006fc}
R.~Britto and B.~Feng,
\newblock Phys. Rev. {\bf D75}, 105006 (2007), hep-ph/0612089.

\bibitem{Badger:2007si}
S.~D. Badger, E.~W.~N. Glover, and K.~Risager,
\newblock JHEP {\bf 07}, 066 (2007), arXiv:0704.3914.

\bibitem{Forde:2007mi}
D.~Forde,
\newblock Phys. Rev. {\bf D75}, 125019 (2007), arXiv:0704.1835.

\bibitem{Ossola:2007ax}
G.~Ossola, C.~G. Papadopoulos, and R.~Pittau,
\newblock JHEP {\bf 03}, 042 (2008), arXiv:0711.3596.

\bibitem{Britto:2007tt}
R.~Britto and B.~Feng,
\newblock JHEP {\bf 02}, 095 (2008), arXiv:0711.4284.

\bibitem{Kilgore:2007qr}
W.~B. Kilgore,
\newblock (2007), arXiv:0711.5015.

\bibitem{Bern:2004ba}
Z.~Bern, D.~Forde, D.~A. Kosower, and P.~Mastrolia,
\newblock Phys. Rev. {\bf D72}, 025006 (2005), hep-ph/0412167.

\bibitem{Dixon:2004za}
L.~J. Dixon, E.~W.~N. Glover, and V.~V. Khoze,
\newblock JHEP {\bf 12}, 015 (2004), hep-th/0411092.

\bibitem{Badger:2004ty}
S.~D. Badger, E.~W.~N. Glover, and V.~V. Khoze,
\newblock JHEP {\bf 03}, 023 (2005), hep-th/0412275.

\bibitem{Badger:2005jv}
S.~D. Badger, E.~W.~N. Glover, and V.~V. Khoze,
\newblock JHEP {\bf 01}, 066 (2006), hep-th/0507161.

\bibitem{Gorsky:2005sf}
A.~Gorsky and A.~Rosly,
\newblock JHEP {\bf 01}, 101 (2006), hep-th/0510111.

\bibitem{Mansfield:2005yd}
P.~Mansfield,
\newblock JHEP {\bf 03}, 037 (2006), hep-th/0511264.

\bibitem{Ettle:2006bw}
J.~H. Ettle and T.~R. Morris,
\newblock JHEP {\bf 08}, 003 (2006), hep-th/0605121.

\bibitem{Ettle:2007qc}
J.~H. Ettle, C.-H. Fu, J.~P. Fudger, P.~R.~W. Mansfield, and T.~R. Morris,
\newblock JHEP {\bf 05}, 011 (2007), hep-th/0703286.

\bibitem{Ettle:2008ey}
J.~H. Ettle, T.~R. Morris, and Z.~Xiao,
\newblock JHEP {\bf 08}, 103 (2008), arXiv:0805.0239.

\bibitem{Ettle:2008ed}
J.~H. Ettle,
\newblock (2008), arXiv:0808.1973.

\bibitem{Mason:2005kn}
L.~J. Mason and D.~Skinner,
\newblock Phys. Lett. {\bf B636}, 60 (2006), hep-th/0510262.

\bibitem{Mason:2005zm}
L.~J. Mason,
\newblock JHEP {\bf 10}, 009 (2005), hep-th/0507269.

\bibitem{Boels:2007qn}
R.~Boels, L.~Mason, and D.~Skinner,
\newblock Phys. Lett. {\bf B648}, 90 (2007), hep-th/0702035.

\bibitem{Boels:2006ir}
R.~Boels, L.~Mason, and D.~Skinner,
\newblock JHEP {\bf 02}, 014 (2007), hep-th/0604040.

\bibitem{Boels:2007gv}
R.~Boels,
\newblock Phys. Rev. {\bf D76}, 105027 (2007), arXiv:hep-th/0703080.

\bibitem{Boels:2008fc}
R.~Boels, K.~J. Larsen, N.~A. Obers, and M.~Vonk,
\newblock JHEP {\bf 11}, 015 (2008), arXiv:0808.2598.

\bibitem{Jiang:2008xw}
W.~Jiang,
\newblock (2008), arXiv:0809.0328.

\bibitem{Boels:2007pj}
R.~Boels and C.~Schwinn,
\newblock Phys. Lett. {\bf B662}, 80 (2008), arXiv:0712.3409.

\bibitem{Boels:2008ef}
R.~Boels and C.~Schwinn,
\newblock JHEP {\bf 07}, 007 (2008), arXiv:0805.1197.

\bibitem{Boels:2007qy}
R.~Boels, C.~Schwinn, and S.~Weinzierl,
\newblock PoS {\bf RADCOR2007}, 016 (2007), arXiv:0712.3506.

\end{thebibliography}
\bibliographystyle{/home/stefanw/latex-style/h-physrev5}

\end{document}